\documentclass{aa} 

\usepackage{natbib}
\usepackage{graphicx}
\usepackage{txfonts}
\usepackage{epsfig}
\usepackage{times}
\usepackage{rotating}
\usepackage{float}
\usepackage[caption = false]{subfig}
\usepackage{setspace}
\usepackage{lscape}
\usepackage{epstopdf}
\usepackage{tabularx}
\usepackage{caption}
\usepackage[usenames]{color}
\usepackage{auto-pst-pdf}
\usepackage[toc,page]{appendix}
\usepackage{soul}
\usepackage{amsmath}
\usepackage[]{hyperref}

\newcommand{\HIIphot}{\textsc{\HIIphot}}

\newcommand*{\kpc}{\ifmmode{\rm kpc}\else kpc \fi}
\newcommand*{\Mpc}{\ensuremath{{\rm Mpc}}}
\newcommand*{\Myr}{\ifmmode{\rm Myr}\else Myr \fi}
\newcommand*{\Gyr}{\ifmmode{\rm Gyr}\else Gyr \fi}

\newcommand*{\TNG}{\textsc{TNG}50-1}

\newcommand*{\Msun}{\ensuremath{{\rm M}_{\odot}}}

\newcommand*{\BDO}{\textsc{BD}}
\newcommand*{\DDO}{\textsc{BL}}
\newcommand*{\CS}{\textsc{CS}}

\newcommand*{\BH}{Massive Accretion}
\newcommand*{\BL}{No Massive Accretion}

\newcommand*{\BEARD}{\textsc{BEARD}}

\newcommand*{\LCDM}{\ensuremath{\Lambda}CDM}

\newcommand{\BulgeDomniatedEnum}{0}
\newcommand{\ControlSampleEnum}{1}
\newcommand{\DiskDominatedEnum}{2}

\newcommand*{\StellarMassLowLim}{\ensuremath{1\times10^{10}\,\Msun}}
\newcommand*{\StellarMassUpLim}{\ensuremath{5\times10^{11}\,\Msun}}
\newcommand*{\SubhaloIsolationRadii}{\ensuremath{500~\kpc}}

\newcommand*{\ParentSampleN}{\ensuremath{538}}
\newcommand*{\SubSampleN}{\ensuremath{135}}

\newcommand*{\NSatDisk}{\ensuremath{515}}
\newcommand*{\NSatBulge}{\ensuremath{1051}}
\newcommand*{\NSatControl}{\ensuremath{506 \pm 50}}

\newcommand*{\SatelliteMaxAperture}{\ensuremath{300~\kpc}}
\newcommand*{\SatelliteMinPart}{\ensuremath{50}}
\newcommand*{\SatelliteMinMagnitude}{\ensuremath{-11~{\rm mag}}}

\newcommand*{\BtoD}{bulge-to-disc}

\newcommand{\lfSlope}[1]{
    \ifcase#1\relax
    \ensuremath{-1.28  \pm 0.01}
    \or 
    \ensuremath{-1.27 \pm 0.02}
    \or
    \ensuremath{-1.35 \pm 0.02}
    \fi
}

\newcommand{\lfBreakMag}[1]{
    \ifcase#1\relax
    \ensuremath{-22.0  \pm 0.3} 
    \or 
    \ensuremath{-21.8 \pm 0.4}
    \or
    \ensuremath{-21.9 \pm 0.6} 
    \fi
}

\newcommand{\lflogPhi}[1]{
    \ifcase#1\relax
    \ensuremath{-5.49 \pm 0.07} 
    \or 
    \ensuremath{-5.77 \pm 0.10}
    \or
    \ensuremath{-6.01 \pm 0.14} 
    \fi
}

\newcommand{\rNFWFitLogc}[1]{
    \ifcase#1\relax
    \ensuremath{0.25\pm0.05} 
    \or 
    \ensuremath{0.48\pm0.06} 
    \or
    \ensuremath{0.79\pm0.05} 
    \fi
}

\newcommand{\NsatScalling}[1]{
    \ifcase#1\relax
    \ensuremath{\beta_{0} = 0.53_{-0.10}^{+0.09}}
    \or
    \ensuremath{\beta_{1} = 0.74_{-0.07}^{+0.07}}
    \or
    \ensuremath{\beta_{2} = 0.15_{-0.23}^{+0.23}}
    \or 
    \ensuremath{\alpha = 9.5_{-3.3}^{+3.4}}
    \fi
}

\newcommand{\NsatScallingSimple}[1]{
    \ifcase#1\relax
    \ensuremath{\beta_{0} =  0.52_{-0.10}^{+0.09}}
    \or
    \ensuremath{\beta_{1} = 0.76_{-0.05}^{+0.05}}
    \or
    \ensuremath{\alpha = 9.4_{-3.0}^{+3.3}}
    \fi
}

\newcommand{\NsatScallingPoisson}[1]{
    \ifcase#1\relax
    \ensuremath{\beta_{0} =   0.51_{-0.08}^{+0.08}}
    \or
    \ensuremath{\beta_{1} =  0.74_{-0.05}^{+0.05}}
    \or
    \ensuremath{\beta_{2} = 0.20_{-0.15}^{+0.15}}
    \fi
}

\newcommand{\NsatScallingPoissonSimple}[1]{
    \ifcase#1\relax
    \ensuremath{\beta_{0} =   0.51_{-0.08}^{+0.08}}
    \or
    \ensuremath{\beta_{1} =  0.77_{-0.04}^{+0.04}}
    \fi
}

\newcommand{\AngularAlignement}[1]{
    \ifcase#1\relax
    \ensuremath{\sim-2.3~{\rm deg}/\Gyr} 
    \or 
    \ensuremath{\sim-1.5~{\rm deg}/\Gyr}
    \or
    \ensuremath{\sim-4.5~{\rm deg}/\Gyr}
    \fi
}

\usepackage{xparse}
\usepackage[normalem]{ulem}

\NewDocumentCommand{\updated}{m o}{%
  \IfNoValueTF{#2}
    {
      \ifmmode
        \mathbf{#1}%
      \else
        \textbf{#1}%
      \fi
    }
    {
      \ifmmode
        \text{\sout{#2}}%
        \mathbf{#1}%
        
      \else
        \sout{#2}\,\textbf{#1}%
      \fi
    }%
}

\begin{document}

\title{Bulgeless Evolution And the Rise of Discs (BEARD)}
\subtitle{III. A numerical simulation view of satellites around Milky-Way analogues.}

\authorrunning{S. Cardona-Barrero et al.}

\titlerunning{BEARD III. }

\author{
Salvador Cardona-Barrero\inst{1,2},
Jairo Méndez-Abreu\inst{1,2},
Adriana de Lorenzo-Cáceres\inst{1,2},
Carlos Marrero de la Rosa\inst{2,1},
Yetli Rosas-Guevara\inst{3, 14},
Elena Arjona-Gálvez\inst{2,1},
Mario Chamorro Cazorla\inst{4, 5},
Nelvy Choque-Challapa\inst{6},
Enrico Maria Corsini\inst{7, 8}
Arianna Di Cintio\inst{1, 2},
David Fernandez\inst{9,10,11},
Daniele Gasparri\inst{12},
Divakara Mayya\inst{9}, 
Lorenzo Morelli\inst{12},
Francesca Pinna\inst{2,1},
Alessandro Pizzella\inst{7, 8}, 
Javier Román\inst{4, 14},
Daniel Rosa Gonzalez\inst{9}, 
Olga Vega\inst{9} \and
Stefano Zarattini\inst{13} 
}

\institute{
Departamento de Astrof\'isica, Universidad de La Laguna, Avenida Astrof\'isico Francisco S\'anchez s/n, E-38206 La Laguna, Spain.
\and
Instituto de Astrof\'isica de Canarias, calle Vía L\'actea s/n, E-38205 La Laguna, Tenerife, Spain. 
\and 
Donostia International Physics Centre (DIPC), Paseo Manuel de Lardizabal 4, 20018 Donostia-San Sebastian, Spain. 
\and
Departamento de Física de la Tierra y Astrofísica, Universidad Complutense de Madrid, E-28040 Madrid, Spain.
\and
Instituto de Física de Partículas y del Cosmos (IPARCOS), Facultad de Ciencias Físicas, Universidad Complutense de Madrid, E-28040 Madrid, Spain. 
\and 
Departamento de Física, Universidad Técnica Federico Santa Maria, Av. Vicuña Mackenna 3939, 8940897, San Joaquín, Santiago, Chile. 
\and
Dipartimento di Fisica e Astronomia “G. Galilei”, Università di Padova, vicolo dell’Osservatorio 3, I-35122 Padova, Italy.
\and
INAF-Osservatorio Astronomico di Padova, vicolo dell’Osservatorio 5, I-35122 Padova, Italy 
\and
Instituto Nacional de Astrofísica, Óptica y Electrónica, Luis Enrique Erro 1, Tonantzintla, 72840, Puebla, Mexico.
\and
Planetarium La Enseñanza, Medellín, Antioquia, CP. 050022 Colombia.
\and
Canada–France–Hawaii Telescope, Kamuela, HI 96743, USA.
\and
Instituto de Astronomía y Ciencias Planetarias, Universidad de Atacama, Copayapu 485, Copiapó, Chile.
\and
Centro de Estudios de Física del Cosmos de Aragón (CEFCA), Plaza San Juan 1, 44001 Teruel, Spain.
\and 
Departamento de F\'isica, Universidad de C\'ordoba, Campus Universitario de Rabanales, Ctra. N-IV Km. 396, E-14071 C\'ordoba, Spain 
\\
}

\date{\today}

\abstract{}
{
 The existence of massive disc galaxies with little or no bulge challenges conventional $\Lambda$ cold dark matter model, which typically favours dynamically hot central structures due to early collapse and mergers. The study of these bulgeless disc galaxies is the aim of the Bulgeless Evolution And the Rise of Discs (BEARD) survey, as they offer a unique opportunity to investigate the link between galaxy morphology and the properties of their satellite systems.
}
{
 Using the high-resolution cosmological hydrodynamical simulation \TNG{}, we studied the satellite populations of \SubSampleN{} bulgeless galaxies. We compared their satellite properties to those of a bulge-dominated control sample with matched stellar masses. 
 Our analysis focuses on satellite abundance, luminosity functions, spatial distribution, orbital alignment, and infall histories.
}
{
 We find that satellite abundance is largely independent of host galaxy morphology. 
 However, satellites around bulgeless galaxies  exhibit luminosity functions with a steeper faint-end slope, are more centrally concentrated, and show stronger orbital alignment with the host disc plane.
  The orbital alignment originates from coherent post-infall dynamical evolution that depends on host galaxy morphology. The infall of more massive satellites can additionally perturb this process, contributing to a weakening or temporary stalling of the secular alignment.
}
{
 Due to the co-evolution of the host galaxy and the satellite system, the morphology of the central galaxy leaves a clear imprint on its satellite system.
Bulgeless galaxies tend to have dynamically colder, more aligned, and more centrally concentrated satellite populations. These trends reflect a more quiet merger history and support the use of satellite properties as tracers of host galaxy formation pathways.
}

\maketitle


\section{Introduction}
\label{sec:intro}


The stochastic nature of galaxy growth within the $\Lambda$ cold dark matter (\LCDM{}) framework favours the formation of dynamically hot central structures, such as classical bulges. 
These components are typically thought to arise either from early gravitational collapse or through the hierarchical merging of smaller progenitors at high redshift \citep{DeLucia2011,Brooks2016,Costantin2021,Costantin2022}. 

Within this context, the existence of massive disc galaxies ($M_{\ast} > 10^{10}\,\Msun$) with little or no bulge remains a puzzle \citep{Kautsch2009, Barazza2009, Kormendy2010, DiTeodoro2023}. 
The specific formation pathways of these bulgeless disc galaxies shape not only their internal structure and morphology \citep[e.g.][]{Sales2012, Du2021, Yetli} but also influence the properties of their satellite systems.

In the dwarf regime, galaxy formation models and subgrid physics prescriptions are often calibrated using the observed properties of Local Group dwarfs, particularly the satellite galaxies of the Milky Way (MW). 
Thanks to its proximity, the MW offers a uniquely detailed view of satellite populations \citep[e.g. ][]{McConnachie2012,Simon2019}. 
Historically, the abundance of satellites on the MW has been at the heart of one of the most well-known challenges to the \LCDM{} model, the missing satellites problem \citep{Klypin1999,Moore1999}. 
Over time, however, this issue has become better understood. Factors such as subhalo occupation statistics, galaxy formation efficiency, and inclusion of baryonic processes in simulations have all contributed to alleviating, or even solving, the tension originally raised by the discrepancy between DM only simulations and early observations of the MW satellite population 
(see \citealt{Sales2022, Jung2024} and references there in).

Since satellites and their hosts co-evolve over extended periods \citep{Wetzel2015ELVIS, Simpson2018}, satellite populations may carry imprints of their host's formation history, morphology, and evolution. 
Their global properties, such as luminosity functions and radial distributions, are sensitive to the physics of reionisation \citep[e.g.][]{Bullock2000,Okamoto2008}, environmental effects such as tidal and ram-pressure stripping \citep[e.g.][]{Mayer2006,Emerick2016}, and the structural properties of the host itself \citep[e.g.][]{GarrisonKimmel2017}.

It also remains unclear whether the MW satellite system is representative of galaxies of similar mass in the Universe or whether it is an outlier. Cosmological simulations suggest that the MW satellite population may be atypical in several respects: The presence of two massive satellites is uncommon \citep{Haslbauer2024,Buch2024}, and the radial distribution of its satellites is more centrally concentrated than expected \citep[e.g.][]{Carlsten2020}.
The MW also exhibits an unusually high quenched fraction, nearly all of its satellites are quenched, exceeding that observed around comparable hosts in surveys such as SAGA and ELVES \citep{Karunakaran2023,Geha2024}, possibly reflecting the MW’s relatively quiet merger history \citep{Kruijssen2019}. High-resolution simulations similarly show that MW–like halos can display elevated quenched fractions \citep[e.g.][]{Engler2023,Rodriguez-Cardoso2025}, although such results fall within a broader halo-to-halo scatter.
Both simulations and observations consistently indicate substantial system-to-system variation in satellite population properties, even at a fixed host mass \citep{Font2021,Engler2023,Mao2024}, underscoring the need for large statistical samples when assessing how typical the MW truly is.

Over the past years, large observational programs such as ELVES \citep{Carlsten2022ELVES}, SAGA \citep{Geha2017}, xSAGA \citep{Wu2022xSAGA}, and others \citep[e.g.][]{Zaritsky2024} have significantly expanded the sample of galaxies with known satellite populations. 
These efforts have collectively enabled statistical analyses over samples of tens to hundreds of nearby galaxies, improving our understanding of satellite demographics beyond the Local Group. 
However, most of these works have focused on general MW-mass hosts, often without isolating the impact of bulge morphology.

The Bulgeless Evolution And the Rise of Discs\footnote{\url{http://www.beardsurvey.com}} (\BEARD{}) survey  (Méndez-Abreu in prep.), is an observational program targeting a volume-limited ($<40~\Mpc$) sample of $75$ galaxies, of which $54$ are massive bulgeless spirals ($M_{\ast} \ge 10^{10}\Msun$). Its main scientific aim is to recover the past merger history and mass growth of bulgeless galaxies. 
Bulgeless galaxies are defined here as systems with a bulge-to-disc mass ratio $<0.08$ and a bulge-to-total mass ratio $<0.1$, following the photometric decomposition of Zarattini (in prep.).

Understanding whether such galaxies pose a challenge to \LCDM{} theories is therefore of central importance. The \BEARD{} project is a multi-facility survey, and among its different observations, we carried out a deep photometric survey covering $100~\kpc$ radii around our bulgeless galaxies. The \BEARD{} large field of view and photometric depth enable simultaneous characterisation of both the central galaxies \citep{Carlos} and their satellite populations  (Choque-Challapa in prep.).\footnote{Satellite galaxies are not uniformly distributed within the virial radius, but tend to be concentrated close to its host galaxy. Data from MW analogues \citep[e.g.][]{Mao2024} and from simulation of MW-like systems (see Sec.~\ref{sec:radial} and references there in) indicate that within the \BEARD{} aperture we expect to capture between $20\%$ and $70\%$ of the full satellite population of each host.}

In this work, we use the cosmological hydrodynamical simulation \TNG{} \citep{Pillepich2018TNG, Nelson2018TNG} to study the satellite populations of massive bulgeless galaxies, and to investigate how their properties correlate with host morphology.
This paper is structured as follows: 
In Sec.~\ref{sec:methods} we describe the simulations and galaxy identification. 
In Sec.~\ref{sec:selection}  we present the sample of host bulgeless galaxies, we describe the construction of a comparison  control sample and the satellite selection criteria. 
In Sec.~\ref{sec:sat-pop} we explore the properties of the satellite galaxies at a population level, comparing across the different samples.
Finally, in Sec.~\ref{sec:summary} we provide a summary of our findings together with our conclusions.

\section{Methodology}
\label{sec:methods}
\subsection{The TNG50 simulation}
\label{sec:tng50}

For this study, we used the \TNG{} simulation \citep{Pillepich2019TNG50, Nelson2019TNG50} from the IllustrisTNG project \citep{Pillepich2018TNG, Marinacci2018TNG, Naiman2018TNG, Springel2018TNG, Nelson2018TNG, Nelson2019DataRelease}. 
IllustrisTNG\footnote{\url{https://www.tng-project.org/}} is a suite of cosmological magneto-hydrodynamic simulations that follow the evolution of structure formation in a $\Lambda$CDM Universe using the moving mesh code \textsc{AREPO} \citep{Springel2010AREPO}. 
The simulations assume a cosmology consistent with results from the Planck Collaboration \citep{Planck2016}, specifically: $h = 0.6774$, $\Omega_{\mathrm{m}} = 0.3089$, $\Omega_{\Lambda} = 0.6911$, and $\sigma_8 = 0.8159$. 
Initial conditions were generated at redshift $z = 127$ and evolved forward to $z = 0$.

Among the different realisations of the IllustrisTNG project, we focused on the highest-resolution run, \TNG{} \citep{Pillepich2019TNG50}. This simulation follows the co-evolution of dark matter, gas, stars, and supermassive black holes within a $(51.7\,\mathrm{Mpc})^3$ comoving volume, using $2160^3$ dark matter particles and an equal number of initial gas resolution elements.

The dark matter particle mass is $m_{\mathrm{DM}} = 4.5 \times 10^5\,\mathrm{M}_\odot$, while the mean baryonic mass resolution is $m_{\mathrm{baryon}} = 8.5 \times 10^4\,\mathrm{M}_\odot$. Stellar particles inherit similar initial masses and lose a fraction of their mass over time due to stellar evolution.

The gravitational softening length for collisionless particles (dark matter and stars) is $\epsilon = 575\,{\rm comoving}\, \mathrm{pc}$ until $z = 1$, after which it is fixed at $\epsilon = 288\,\mathrm{pc}$ in physical units. Gas cells have adaptive resolution, reaching sizes as small as $\sim 72\,\mathrm{pc}$ in dense, star-forming regions.

TNG50 incorporates detailed subgrid physics to model unresolved processes, including radiative cooling, chemical enrichment, and a multiphase interstellar medium with stochastic star formation \citep{Pillepich2018}. Stellar feedback is implemented via kinetic galactic winds, while black hole feedback follows a dual-mode model: thermal at high accretion rates and kinetic at low accretion rates \citep{Weinberger2017, Pillepich2018}.
Thanks to its combination of high resolution and  cosmological framework, \TNG{} is particularly well suited for investigating the demographics and structural properties of satellite galaxies within realistic galactic environments.

\subsection{Halo and subhalo identification}

All halo and merger tree data used in this work were obtained from the publicly released \TNG{} catalogues \citep{Nelson2019DataRelease}, where a complete description can be found. Here, we provide a brief summary.

Dark matter halos and subhalos in the TNG simulations were identified using the \textsc{Subfind} algorithm \citep{Springel2001SubFind}, applied to friends-of-friends (FoF) groups constructed from the dark matter distribution with a standard linking length of $b=0.2$ times the mean inter-particle separation. 
The \textsc{FoF} algorithm identifies virialised structures (halos), while \textsc{Subfind} subsequently decomposes each halo into gravitationally self-bound substructures (subhalos), including both central and satellite galaxies.

Each subhalo is assigned a host FoF group, and baryonic components (gas, stars, black holes) are associated with the closest bound dark matter particle. The most massive subhalo in a given FoF group is considered the central, while the others are labelled as satellites. 
This inclusive halo-finding approach makes it possible to track galaxies within the potential wells of larger systems; however, we note that subhalo identification remains a complex challenge. Recent studies (e.g. \citealt{Mansfield2024,Diemer2024,Forouhar2025}) have shown that many halo finders can struggle to recover subhalos after pericentric passages, when there is substantial mass loss and blending with the central host density, making their detection more difficult.

To study the evolutionary history of halos and galaxies, we used merger trees constructed via the \textsc{Sublink} algorithm \citep{Rodriguez2015}. \textsc{Sublink} links subhalos across snapshots by tracing the most bound particles, enabling the reconstruction of the formation and accretion history of both centrals and satellites.

\section{Sample selection}
\label{sec:selection}

\subsection{Host galaxies}
\label{sec:selection-hosts}

\begin{figure}
    \centering
    \includegraphics[width=1\linewidth]{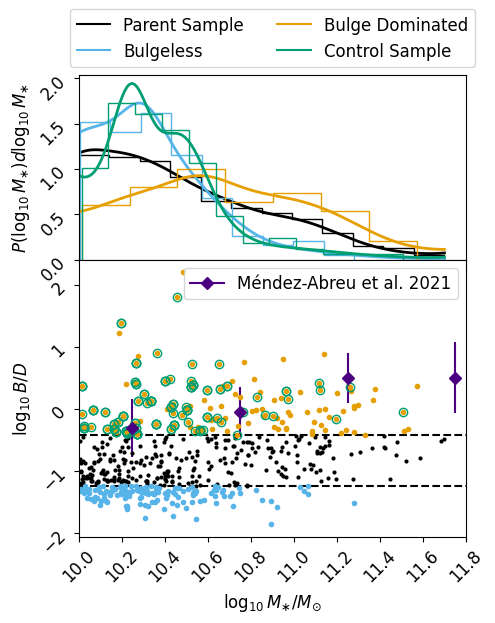}
    \caption{
    Top panel: Distribution of host galaxy stellar mass. 
    Bottom panel: Bulge-to-disc (B/D) mass ratio of host galaxies as a function of stellar mass. 
    Indigo diamonds show the relation from \cite{Mendez-Abreu2021}, obtained under the assumption $B+D=T$ (see Section~\ref{sec:selection-hosts}). \\
    Both panels display the full parent sample (black), \DDO{} sample (blue), the \BDO{} sample (orange), and, for clarity, one realisation of the \CS{} (green).
    In the bottom panel, black horizontal dashed lines mark the first and fourth quantiles of the \BtoD{} distribution, which define the division between \DDO{} and \BDO{} galaxies.
    }
    \label{fig:BD}
\end{figure}

We are interested in exploring the properties of the satellite population of \BEARD-like 
galaxies. 
Thus, from the \TNG{} simulation we identified all the central subhalos whose stellar mass is between \StellarMassLowLim{} and \StellarMassUpLim{}. 
Close galaxy pairs would indeed affect our measurements of the satellite populations, thus we decided to include an isolation criterion: We excluded from the sample any host galaxy that has a companion within \SubhaloIsolationRadii{} whose stellar mass is greater than $80\%$ of the candidate's.
This selection resulted in a parent sample of \ParentSampleN{}  galaxies. 

\begin{table}[]
    \centering
    \caption{Brief description of the different datasets. }
    \begin{tabular}{|c|cc|}
    \hline 
         &  \# of Hosts & \# of Satellites \\
    \hline\hline
         \DDO{} & \SubSampleN{} & \NSatDisk{} \\
         \CS{} & \SubSampleN{} & \NSatControl{}\\
        \BDO{} & \SubSampleN{} & \NSatBulge{} \\
     \hline 
    \end{tabular}
    \tablefoot{Number of host and satellite galaxies (first and second column respectively). The number of satellites of the \CS{} sample indicates the mean and $1\sigma$ scatter over the $1000$ realisations.}
    \label{tab:datasets}
\end{table}

\begin{figure*}
    \centering
    \includegraphics[width=1\linewidth]{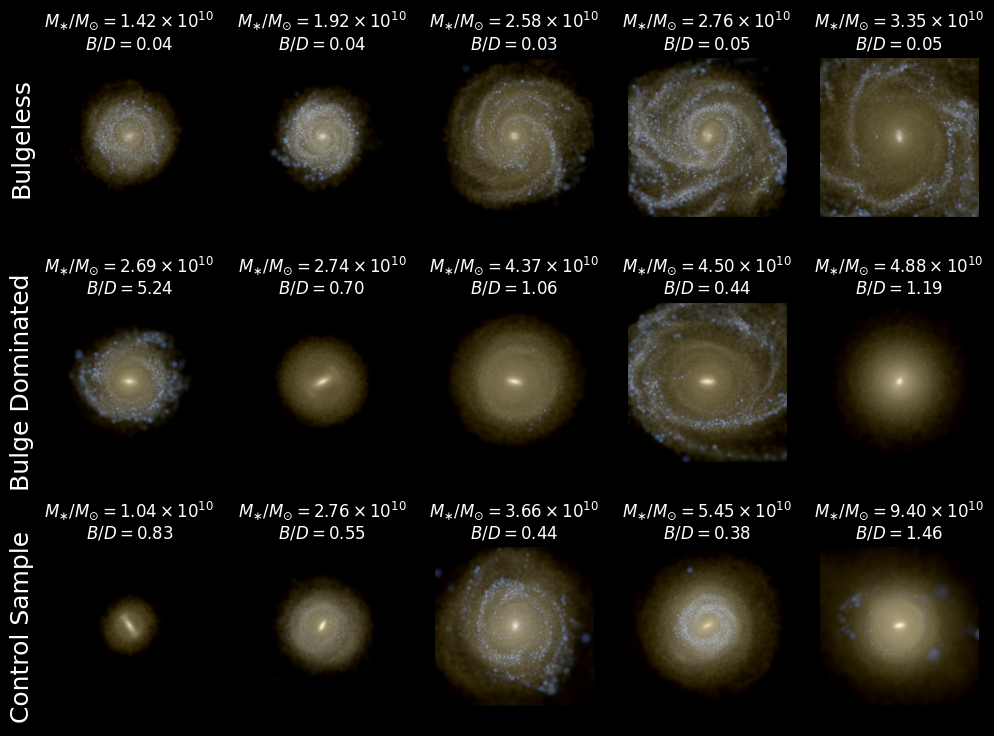}
    \caption{Face-on color-composite images of randomly selected galaxies from the samples used in this work. 
    Images have been generated with the python module \texttt{pynbody}\citep{pynbody}.
    The three channels are the stellar surface brightness in the  $i$, $v$, and $u$ Johnston filters.       
    From top to bottom: \DDO{}, \BDO{}, and \CS{} galaxies. 
    From left to right, galaxies are sorted by stellar mass in increasing order. 
    Each image has a width of $50\,\kpc$. 
    Stellar mass and $B/D$ is given for each galaxy.
    }
    \label{fig:host-renders}
\end{figure*}

In addition, we imposed a morphological constraint to select only bulgeless galaxies. To this aim, we split the parent sample according to its \BtoD{} mass ratio ($B/D$).
We obtained the \BtoD{} mass ratio from the kinematic decomposition of \cite{Zana2022}. Here, we provide a brief description of their method, but we refer the interested reader to the original work for further details.
The decomposition is based on analysing the binding energy and circularity ($\eta\equiv j_z/j_{\rm circ}(E)$ with $j_z$, the angular momentum perpendicular to the plane, and $j_{\rm circ}$, the angular momentum of a circular orbit with energy $E$) distribution of stellar particles to classify them into five main structures: thin disc, thick disc, pseudo-bulge, bulge, and stellar halo.
The classification begins by separating stellar particles into more and less bound components based on their binding energy. A Monte Carlo sampling technique is then applied to the most bound sample, assigning particles to the bulge and selecting those with $\eta < 0$ along with their symmetric positive counterparts\footnote{This is done for the classical bulge component to have approximately zero net rotation, if only stars with negative circularity were selected the bulge component would have a non zero negative angular momentum.}. The remaining particles are classified as part of the thin disc if their circularity is sufficiently high ($\eta > 0.7$) or as part of the pseudo-bulge.
A similar approach is used for the less bound component to determine whether particles belong to the stellar halo, thin disc, or thick disc.

This procedure does not take into account the possible existence of a bar. The star particles forming a bar (if any) are likely assigned to the pseudo-bulge component, with some contamination in all the other ones.

This kinematic decomposition is not likely to be a one-to-one  comparison with  photometric decompositions \citep[see the discussion in][]{cristiani_untangling_2024} such as the one used in the \BEARD{} survey (Zarattini in prep.). Thus, a hard cut in \BtoD{} mass ratio equal to the one used in observational studies may not be adequate. 
In contrast we decide to get as our sample of BulgeLess (\BEARD-Like, \DDO{})  galaxies those that lie in the first quartile of the \BtoD{} mass ratio distribution (Fig. \ref{fig:BD}).  
Note that in this work the disc component includes thin and thick disc, while the bulge component only includes the classical bulge. 
With this approach, we effectively select the most extreme \DDO{} galaxies in a volume limited sample. 
Moreover, we  select as \DDO{} galaxies those whose bulge mass is only as much as $\lesssim 6\%$ per-cent of the disc mass, close to the $8\%$  observed for our MW \citep{Shen2010}.

To understand the properties and particularities of the satellite population of \DDO{} galaxies, we need a comparison sample. For doing so, we decide to compare with bulge dominated (\BDO{}) galaxies (the fourth quartile in the \BtoD{} mass ratio distribution, $B/D \gtrsim 0.38$). 

However, as expected from previous works \citep[e.g.][]{Weinzirl2009,Mendez-Abreu2021}, it is clear from Fig.~\ref{fig:BD} that there is a mass dependence of the \BtoD{} mass ratio. 
In the original study from \cite{Mendez-Abreu2021}, the authors report a positive correlation between the bulge-to-total mass ratio ($B/T$) and stellar mass using data from the  CALIFA IFS survey \citep{Sanchez2016califaifs}.
Here, in Fig.~\ref{fig:BD}, we have transformed their data by assuming a two component galaxy (only bulge and disc, $B+D=T$), and thus the \BtoD{} mass ratio can be expressed as $B/D = (B/T)/(1-(B/T))$. This transformation is fully consistent with their formalism, as they assume a two-component galaxy model.

As the number of satellite galaxies is expected to scale with the host halo mass \citep[e.g. ][]{Gao2004,Klypin2011}, and therefore with stellar mass, the mass dependence of the \BtoD{} mass ratio introduces a bias in the comparison of the satellite population properties of the \DDO{} and \BDO{} samples.
In order to overcome this issue, we produced a third sample whose stellar mass distribution is statistically similar to the stellar mass distribution of the \DDO{} sample. We refer to this new sample as control sample (\CS{}). 
In Fig.~\ref{fig:host-renders} we show a RGB composition of the stellar component of a random selection of host galaxies belonging to each of the three samples. In this mosaic the morphological difference between the samples and  massive discs of the \DDO{} sample is clearly visualised.

The \CS{} has been defined as follows. 
First, we estimate the probability distribution of the (log$_{10}$) stellar mass of the \DDO{} $(P_{BL}(\log_{10}M_{\ast}))$ and \BDO{} $(P_{BD}(\log_{10}M_{\ast}))$ samples via kernel density estimation (solid lines in top panel of Fig.~\ref{fig:BD}). 
Then we sample, with replacement, $N$ galaxies (form the \BDO{} sample) with a sampling probability  proportional to $P_{BL}(\log_{10}M_{\ast}) / P_{BD}(\log_{10} M_{\ast})$. 
In  this case $N=\SubSampleN{}$, in order to have the same number of host galaxies in all the samples of the analysis. 
This has the drawback that the same galaxy may be selected more than once. 
However, the method ensures that too massive galaxies, which are overrepresented in the \BDO{} (compared to the \DDO{}) sample, have low sampling probabilities, while those galaxies that are under-represented (i.e. low mass galaxies) have high sampling probabilities. 

The resulting \CS{} has a (log$_{10}$) stellar mass distribution indistinguishable from the \DDO{} sample. 
In Fig.~\ref{fig:BD} we show one realisation of the \CS{} compared to the \DDO{}.
In this work, except when stated otherwise, confidence regions regarding the \CS{}  were obtained by repeated measurements over $1000$ different realisations of the \CS{}.
This prevents us from being biased towards any particular realisation of the \CS{}.
A Kolmogorov–Smirnov (KS) test between the two samples returns a median maximum distance of $0.12$ with a median p-value of $0.3$, not being able to distinguish between the two populations, as expected as the two samples are equal by construction. 
Despite the fact that our selection does not, a priori, enforce matching total halo masses, we have verified that the halo mass distributions of the \DDO{} and \CS{} samples are statistically consistent (see appendix~\ref{apx:halomass}), and that their mass accretion histories are also nearly identical (see appendix~\ref{apx:MAH}). This confirms that the results presented here are not influenced by differences in the underlying host properties.

\subsection{Satellite galaxies}
\label{sec:selection-satellites}

We identified as satellite every subhalo located within a $3$D aperture of \SatelliteMaxAperture{} from the host. 
This aperture is approximately the virial radius of a MW-like halo\footnote{In this context, the virial radius is defined as the spherical aperture that encloses a mean density $\Delta_{\rm vir}$ times the critical density of the Universe. The value of $\Delta_{\rm vir}$ can be derived from the spherical collapse model and depends on the cosmology and redshift through the matter density parameter $\Omega_m(z)$. A commonly used fitting formula \citep{Bryan1998} gives $\Delta_{\rm vir} = 18\pi^2 + 82x - 39x^2, \quad \text{where } x = \Omega_m(z) - 1$. 
Note, however, that in this work for the size estimation of the simulated halos we employed a fixed over density criteria, i.e. $R_{200}$ indicates the radii that the enclosed density is $200$ times the mean density of the Universe.} 
and has been extensively used in the literature for satellite identification 
in both simulations \citep[e.g.][]{samuel2020profile, Engler2021, Font2022} and observations \citep[e.g.][]{Ruiz2015, Mao2021, Mao2024, Carlsten2022ELVES}.
Thus, for consistency with previous works, we decided to fix this aperture in the search of satellite galaxies.

We also imposed some refinement criteria by removing all the subhalos that do not have cosmological origin, i.e. those when traced back in time were formed inside another halo (this has been done according to the \texttt{SubhaloFlag} in the public data release). 
Furthermore, we removed all the subhalos with fewer than \SatelliteMinPart{} star particles, to ensure we are not affected by resolution issues. 
Finally, we mimicked \BEARD{}'s expected magnitude limit by imposing a further cut in the r band absolute magnitude removing all satellites dimmer than \SatelliteMinMagnitude{} in the $r$ band (Choque-Challapa in prep.).
This selection implies an effective lower stellar mass cut of the satellite galaxies of $\sim2\times10^6\,\Msun{}$\footnote{In \TNG{}, resolution effects in the satellite stellar to halo mass relation emerge below stellar masses of $10^6\Msun{}$ (see Appendix A of \citealt{Engler2021}). With our selection we do not suffer any possible completeness issue due to resolution effects. }. 
As we are only focused in massive well resolved satellites we do not include any correction from orphan subhalos. 

This procedure lead to a total of \NSatDisk{} satellites in  the \DDO{} samples, \NSatBulge{} in the \BDO{} sample, and an average  of \NSatControl{} in the \CS{} (Table~\ref{tab:datasets}). 
The strong difference in the satellite number counts between the \DDO{} and \BDO{} samples highlights the need of the mass resampling described above (Sec.~\ref{sec:selection-hosts}).

For comparison throughout this work, we use the catalogue of Local Group dwarfs from \cite{LG_dwarfs}
\footnote{We used version v1.0.6 of the \cite{LG_dwarfs} database, the most recent release at the time of this work, available at \url{https://github.com/apace7/local_volume_database}}. We applied selection criteria consistent with those used for the TNG50 satellite galaxies: We included only galaxies located within $300~\kpc$ of their host and with stellar masses greater than $2\times10^{6}~\Msun$.

\section{Results and discussion}
\label{sec:sat-pop}

\subsection{Satellite number counts}
\label{sec:number-counts}

\begin{figure}
    \centering
    \includegraphics[width=1\linewidth]{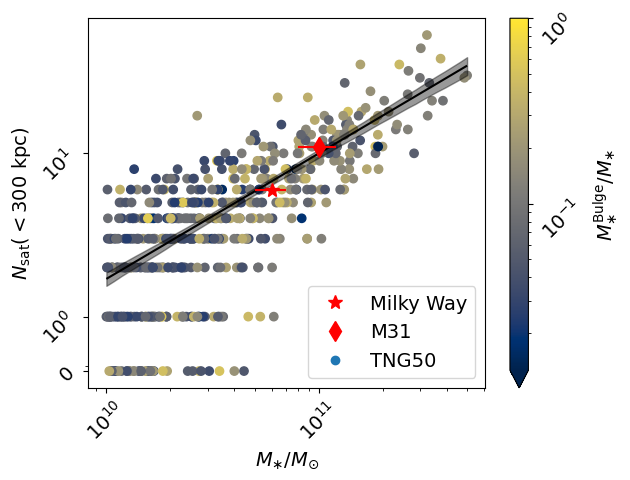}
    \caption{Number of satellite galaxies as function of the host's galaxy stellar mass colour-coded by the host galaxy bulge-to-total mass ratio ($B/T$).
    Note that the y-axis switches from linear to logarithmic scaling at $N_{\rm sat} = 1$.
    The black line shows a power law fit, while the shaded region indicates the $95\%$ confidence interval, estimated from the NB regression. 
    For comparison we include the values of the MW and Andromeda galaxies. The stellar masses have been obtained from \cite{Licquia2015} and \cite{Sick2015} respectively.
    The satellite number counts were obtained from the \cite{LG_dwarfs} database, mimicking our satellite selection criteria (see section~\ref{sec:selection-satellites}).
    }
    \label{fig:nsat-mstar}
\end{figure}

The abundance of satellite galaxies remains sensitive to the physics implemented in simulations. 
While modern cosmological simulations reproduce the observed number of MW satellites \cite[e.g.][]{Engler2021} and the missing satellite problem has largely diminished \citep{Sales2022}; some authors even consider it solved \citep{Jung2024}. Nevertheless, it remains important to examine how satellite number counts scale with host galaxy properties. 
Here, we focus on the relation between the number of satellites and host stellar mass (Fig.~\ref{fig:nsat-mstar}).

We modelled the number of satellites, $N_{\rm sat}$, as a power-law function of the host stellar mass, $M_{\ast}$, using the following form:

\begin{equation}
\label{eq:simple}
    \ln\left(\mathbb{E}\left[N_{\rm sat}
    \right]\right) = \beta_{0} + \beta_{1}\ln{\left(\frac{M_{\ast}}{10^{10}\Msun}\right)},
\end{equation}
where, $\beta_0$ and $\beta_1$ are regression coefficients estimated through a generalised linear model (GLM) using a log link function \citep[e.g.][]{McCullaghNelder1989,dobson2018introduction}.

Motivated by previous studies \citep[e.g.][]{Ruiz2015,Muller2023} suggesting that satellite abundance may also correlate with bulge mass fraction ($B/T=M^{\rm Bulge}_{\ast}/M_{\ast}$), we tested a second regression model in which the slope of the power-law depends linearly on $B/T$:

\begin{equation}
\label{eq:interaction}
    \ln\left(\mathbb{E}\left[N_{\rm sat}
    \right]\right) = \beta_{0} + \left(\beta_{1} +\beta_{2}B/T\right)\ln{\left(\frac{M_{\ast}}{10^{10}\Msun}\right)}.
\end{equation}
To perform the regression, we used both Poisson and Negative Binomial (NB) likelihoods. 
Although Poisson regression is the standard approach for count data, our residuals showed evidence of overdispersion, where the variance exceeds the mean. 
In such cases, the Poisson model may underestimate the parameter uncertainties, leading to biased inferences. 
To address this, we adopted the NB likelihood, which includes an extra dispersion parameter, $\alpha$\footnote{The variance of the NB distribution is $\mu^2/\alpha + \mu$, compared to the variance of the Poisson distribution $\mu$, with $\mu$ the mean value.}, which allows for a more accurate estimation of uncertainties.

We fit a series of Bayesian regression models using the \texttt{bambi} \citep{bambi} Python package. The priors and regression results are summarised in Table~\ref{tab:sclaing_relations}. 
For model comparison, we employ the expected log pointwise predictive density (ELPD), estimated via Pareto-smoothed importance sampling leave-one-out cross-validation (PSIS-LOO, \citealt{vehtari2017practical}), as implemented in \texttt{ArviZ} \citep{arviz}.
The NB models clearly outperform their Poisson counterparts. 
Among the NB models, the simpler version (i.e. without dependence on $B/T$) is slightly preferred.

The presence of over-dispersion is expected as the number of satellite galaxies is more tightly correlated with its host dark matter halo mass (e.g. FoF group mass) than with the central stellar mass alone \citep[see][for an analysis on the non-Poissonity of the subhalo number counts]{Jiang2017}, introducing an additional source of scatter in the relation.
If we had ignored over-dispersion (i.e. using Poisson likelihood only), we would have inferred a statistically significant dependence on $B/T$.

Interestingly, our results are in contrast with those of \cite{Muller2023}, who studied \textsc{TNG100} and found a significant dependence on $B/T$. 
The difference between these two works it is likely related to the resolution effects and simulation volume. 
In our analysis we use \textsc{TNG50} because of its higher resolution compared with \textsc{TNG100}. As discussed in Sec.~\ref{sec:selection-satellites}, this higher resolution, together with our restrictive selection criteria, ensures that our results are not affected by the resolution–driven incompleteness that may influence the sample in \cite{Muller2023}. However, the simulation volume of \textsc{TNG50} is eight times smaller, implying that the most massive host galaxies may be under-represented in our sample. This possible incompleteness at the high–mass end is important to bear in mind, since the strongest correlation between bulge mass fraction and satellite number counts reported by \cite{Muller2023} happens to be stronger in the most massive hosts.

\begin{table*}[]
    \centering
    \caption{Results from the GLM regression.}
    \begin{tabular}{|cc|cccc|c|}
    \hline
        \multicolumn{2}{|c|}{Model} & $\beta_{0}$ & $\beta_{1}$ & $\beta_{2}$ & $\alpha$ & $\Delta$ELPD \\
        \hline
        \multicolumn{2}{|c|}{Priors} & \multicolumn{3}{c}{$\mathcal{N}(0, 10^2)$} & HalfCauchy(1)  & \\
        \hline
       NB & 
       (Eq.~\ref{eq:simple}) & \NsatScallingSimple{0} & \NsatScallingSimple{1} & & \NsatScallingSimple{2} & 0 \\
       NB with B/T dependence & (Eq.~\ref{eq:interaction}) & \NsatScalling{0}  & \NsatScalling{1}  &  \NsatScalling{2}  & \NsatScalling{3}  & 0.25 \\
       \hline
       Poisson with B/T dependence & (Eq.~\ref{eq:interaction}) & \NsatScallingPoisson{0} & \NsatScallingPoisson{1} & \NsatScallingPoisson{2} & & 45.06 \\
       Poisson & (Eq.~\ref{eq:simple}) & \NsatScallingPoissonSimple{0} & \NsatScallingPoissonSimple{1} & \NsatScallingPoissonSimple{2} & & 46.76 \\  
       \hline
    \end{tabular}
    \tablefoot{Two first rows using the negative binomial likelihood, while the third and fourth ones use Poisson likelihood. 
    Each column indicates one of the parameters of the fit. 
    The over dispersion parameter of the negative binomial regression is $\alpha$. 
    Models are sorted by the expected log pointwise predictive density (ELPD). 
    Poisson regression can be clearly ruled out. 
    Both negative binomial regression models have similar score according to ELPD, moreover, the interaction term ($\beta_{2}$) is compatible with $0$ within the $95\%$  high density interval of the posterior. }
    \label{tab:sclaing_relations}
\end{table*}

In summary, we do not find a significant correlation between host morphology and satellite abundance.
The apparent difference between the \DDO{} and \BDO{} samples is largely explained by their different stellar masses; once we account for mass effects (i.e. when comparing the \DDO{} sample with the \CS{}), the satellite number counts become comparable (Table~\ref{tab:datasets}).

\subsection{Luminosities and colours}
\label{sec:sat-lum}

Galaxy magnitudes were computed by adding the contribution of all star particles within each subhalo. 
In IllustrisTNG, each star particle represents a single stellar population with a Chabrier IMF \citep{Chabrier2003}; its initial metallicity is inherited from the gas from which it formed. 
Stellar photometry is based on BC03 SSP models \citep{Bruzual2003} with Padova isochrones. The default stellar particle magnitudes are intrinsic (dust-free) and reported  in standard photometric bands (e.g. SDSS $r$ band).

\begin{figure}
    \centering
    \includegraphics[width=1\linewidth]{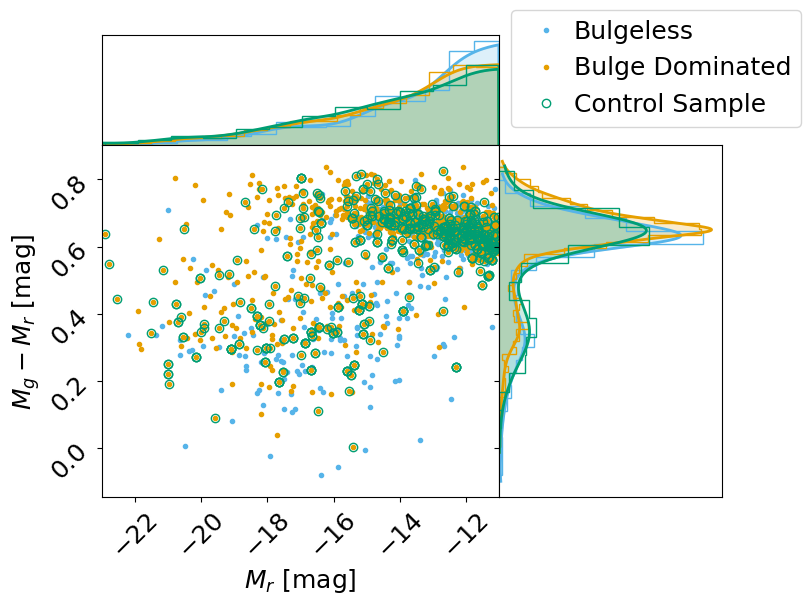}
    \caption{Colour–magnitude diagram of satellite galaxies.
    Satellites from \DDO{} galaxies are in blue, from \BDO{} galaxies in orange, and from the \CS{} in green.
    Each point represents a satellite galaxy in the sample. While the colour distributions are similar across populations, \DDO{} satellites show a noticeable excess at faint magnitudes.
    For visualisation purposes only one realisation of the \CS{} is shown.
    Top and right panels show the $1D$ distribution of r band magnitude and colour respectively. Distributions are shown via KDE (thick lines) and normalised histograms (thin lines).
    }
    \label{fig:ColorMagnitude}
\end{figure}

In Fig.~\ref{fig:ColorMagnitude} we show a colour–magnitude diagram of the satellite galaxies of this study. 
The majority of our satellite galaxies present redder colours, as expected from the large quenched fractions of satellite galaxies in this mass range, which is of the order of $\sim80$ percent \citep[e.g.][]{Donnari2021,Mao2024}.
No clear differences are observed in the overall colour distribution between the different satellite populations; however, an excess of faint satellites is noticeable in the \DDO{} population.

We decided to characterise the differences between satellite populations by estimating their luminosity functions (LFs). For this purpose, we adopted a \cite{Schechter1976} function parametrization:
\begin{equation}
\label{eq:lumfunc}
\begin{split}
n(M; \phi, M^{\rm knee}, \alpha) &= 0.4 \ln(10)\, \phi \, 
10^{0.4 (M^{\rm knee} - M)(\alpha + 1)} \\
&\quad \times \exp\Big[-10^{0.4 (M^{\rm knee} - M)}\Big],
\end{split}
\end{equation}
where $M$ is the satellite absolute magnitude in the $r$ band, $M^{\rm knee}$ is the characteristic magnitude marking the knee of the luminosity function, $\phi$ is the normalisation, and $\alpha$ is the faint-end slope. 

To fit the model to our data, we employed the extended maximum likelihood method introduced in \citet{Andreon2005}, which accounts for variable completeness limits across host galaxies. The corresponding log-likelihood is given by
 
\begin{equation}
    \label{eq:lum-likelihood}
    \ln \mathcal{L}(\{M_{ij}\} \mid \phi, M^{\rm knee}, \alpha) = \sum_{j=1}^{N_{\text{hosts}}} \left( \sum_{i=1}^{N_{\text{sat}}^j} \ln n(M_{ij}; \phi, M^{\rm knee}, \alpha) - s_j \right),
\end{equation}
with the correction term $s_j$ defined as:
\begin{equation}
    s_{j} = \int_{M_{\rm low}^{j}}^{M_{\rm high}^{j}} n(M; \phi, M^{\rm knee}, \alpha)dM.
\end{equation}

Here, we fixed the lower magnitude limit to $M_{\rm low} = -11\,{\rm mag},$, consistent with our satellite selection threshold. The upper limit is fixed at $M_{\rm high} = -100\,{\rm mag}$, following the prescription in \citet{Negri2022}. We verified that this choice does not significantly impact the derived parameters, since bright satellites are rare\footnote{We have also tested defining $M_{\rm high}^{j} = M_{\rm host}^{j}$ for each system individually. This improves constraints on the faint-end slope $\alpha$, but leads to a substantial loss of constraining power in $M^{\rm knee}$ and $\phi$.
}.

We adopted wide uninformative priors for the Schechter function parameters: the faint-end slope $\alpha$, characteristic magnitude $ M^{\ast}$, and normalisation $\log_{10} \phi $. Specifically, the priors are uniform over the ranges $-5 < \alpha < 5$, 
$-30< M^{\rm knee}/\,{\rm mag}<0$, 
and 
$-25< \log_{10} \phi/\,{\rm dex} <15$. 
The posterior distribution was sampled using Markov chain Monte Carlo (MCMC) method, as implemented in the \texttt{emcee} Python package \citep{emcee}, which employs an affine-invariant ensemble sampler.

\begin{figure}
    \centering
    \includegraphics[width=1\linewidth]{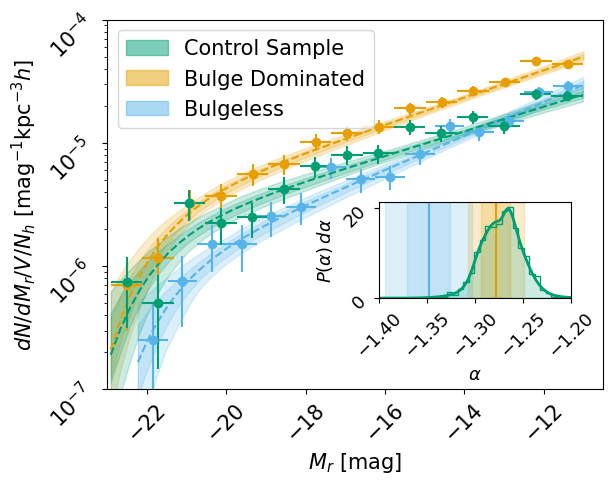}
    \caption{
    Stacked satellite LFs of the \BDO{} (orange), \DDO{} (blue)  and one realisation of the \CS{}. 
    Shaded regions indicate the $68\%$ and $95\%$  confidence intervals.
    Error bars in the data points indicate the $1\sigma$ Poisson uncertainty ($y$-axis) and the bin width ($x$-axis).
    Luminosity functions  have not been fitted to these data-points but to the full distribution (see Sec.~\ref{sec:sat-lum} for details). 
    For visualisation purposes for the \CS{} we show only one random realisation.
    Inset plot: 
    Mean and scatter ($1\sigma$ and $2\sigma$) of the marginalised distribution of the faint end slope for the \DDO{} and \BDO{} samples. 
    For the \CS{} sample we show the distribution of the maximum a posteriori estimates of the faint end slope $\alpha$ over the $1000$ realisations.
    }
    \label{fig:lum-func}
\end{figure}

The resulting luminosity functions are shown in Fig.~\ref{fig:lum-func}.
Posterior estimates of the parameters are summarised in Table~\ref{tab:lf-fit-param}. The \BDO{} sample clearly shows a higher satellite abundance, as seen in the offset of its luminosity function relative to the \DDO{} and \CS{} samples. This emphasises the importance of the \CS{} in distinguishing mass effects from morphological ones.

A visual inspection of the luminosity functions suggests that the \CS{} exhibits an overabundance of bright satellites and a deficit of faint satellites compared to the \DDO{} sample. This trend is more clearly reflected in the marginalised posterior distributions of the faint-end slope $\alpha$ (inset panel of Fig.~\ref{fig:lum-func}), where we find a $\sim 2.8\sigma$ discrepancy: the \DDO{} sample shows a steeper slope, indicating an excess of faint satellites.
Of the three samples, the \DDO{} has a faint-end slope consistent within $1\sigma$ with the MW best-fitting value of $\alpha_{\rm MW} = -1.34^{+0.04}_{-0.03}$ \citep{Nadler2019}, while the other two (\BDO{} and \CS{}) are slightly offset at the $\sim 1.5\sigma$ level (see Table \ref{tab:lf-fit-param}).

The remaining Schechter parameters ($M^{*}$ and $\phi$) are consistent between the \DDO{} and \CS{} samples. This implies that both samples host a similar total number of satellites, in agreement with the results presented in Sec.~\ref{sec:number-counts}, where no significant dependence of satellite abundance on B/T was found. 
However, while the normalisation and characteristic magnitude remain similar, the satellites in \DDO{} systems are systematically fainter. 
 
\begin{figure}
    \centering
    \includegraphics[width=1\linewidth]{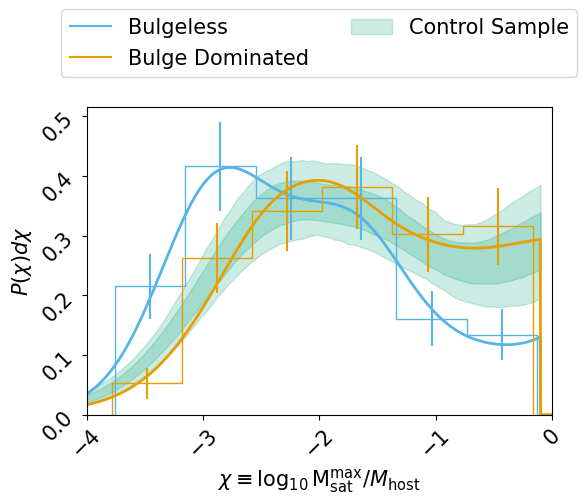}
    \caption{Distribution of the ratio between the stellar mass of the most massive satellite and its host galaxy for the \DDO{} sample (blue), \BDO{} sample (orange), and \CS{} (green). 
    Error bars indicate $1\sigma$ Poisson uncertainty.
    Solid lines are KDE estimations. 
    The green bands encompass the $68\%$ and $95\%$ confidence regions of the estimated PDFs over the 1000 realisations of the \CS{}.
    }
    \label{fig:max-mass}
\end{figure}

Moreover, when examining the stellar mass ratio of the most massive satellite to its host (Fig.~\ref{fig:max-mass}), we see a clear dichotomy that complements with the luminosity function differences. 
There is a  $\sim1~{\rm dex}$ difference between the most likely max-mass ratio between the \DDO{} and \CS{} samples.  
Not only that but \DDO{} systems almost never host a satellite more massive than $\sim10\%$ of its host stellar mass, while both the \BDO{} and \CS{} samples exhibit a significant heavy tail in the most massive satellite mass fraction distribution.

\begin{table}[]
    \centering
    \caption{Luminosity function fit parameters.} 
    \begin{tabular}{|c|ccc|}
    \hline
         &  $\alpha$ & $M^{\rm knee}~[{\rm mag}]$ & $\log_{10}\frac{\phi}{[\kpc^{-3}~h]}$\\
         \hline\hline
        \BDO{} & \lfSlope{\BulgeDomniatedEnum{}} & \lfBreakMag{\BulgeDomniatedEnum{}} & \lflogPhi{\BulgeDomniatedEnum{}}  \\ 
        \CS{}  & \lfSlope{\ControlSampleEnum{}} & \lfBreakMag{\ControlSampleEnum{}} & \lflogPhi{\ControlSampleEnum{}}  \\
       \DDO{} & \lfSlope{\DiskDominatedEnum{}} & \lfBreakMag{\DiskDominatedEnum{}} & \lflogPhi{\DiskDominatedEnum{}} \\
        \hline
    \end{tabular}
    \tablefoot{From left to right: faint end luminosity slope ($\alpha$), knee of the LF ($M^{\rm knee}$), and normalisation ($\phi$).
    Values indicate the mean and $1\sigma$ scatter.
    In the case of the \CS{} the reported value (and error) indicates the average of the maximum a posteriori estimates (and standard deviation) of the $1000$ different realisations.
    }
    \label{tab:lf-fit-param}
\end{table}

\subsection{Satellite spatial distribution}
\label{sec:radial}

\begin{figure}
    \centering
    \includegraphics[width=1\linewidth]{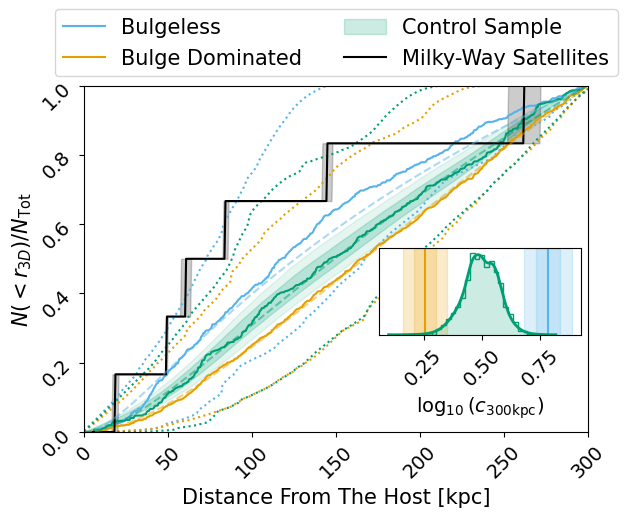}
    \caption{
    Stacked empirical cumulative distribution function of the radial position of the satellites galaxies of the \DDO{} (blue), \BDO{} (orange), and \CS{} (green) samples. 
    The \CS{} shows the $1\sigma$ and $2\sigma$ confidence intervals estimated over $1000$ re-samplings as green bands. 
    The dotted lines indicate the $1\sigma$ scatter due to halo-to-halo variance. 
    Black line shows the distribution of  MW satellites whose positions and errors have been obtained from \cite{LG_dwarfs}. The shadow black region indicates the $1\sigma$ confidence region due to observational uncertainties in the distance measurements.
    Dashed lines indicate the best fit to an NFW radial distribution. 
    The resulting concentrations $c_{300\,\kpc}$ from the fit are shown in the inserted panel in the bottom right of the figure. 
    The MLE of $c_{300\,\kpc}$ for the \DDO{} (blue)  and \BDO{} (orange) samples are shown as a vertical line, the shaded bands indicate $68\%$ and $95\%$ confidence region.
    Regarding the \CS{} we show the distribution of the $1000$ MLE estimations of $c_{300\kpc}$ as a green PDF.
 }
    \label{fig:Sat_rad_dist}
\end{figure}

Another parameter of interest when characterising satellite populations is their spatial distribution, both radial and angular.
In Fig.~\ref{fig:Sat_rad_dist}, we show the radial distribution of the three samples analysed in this work. 

The \DDO{} sample shows a more centrally concentrated satellite distribution compared to both the \BDO{} sample and \CS{}. On average, half of its satellites lie within $\sim120~\kpc{}$, indicating a more compact spatial configuration relative to the other two samples (see Table~\ref{tab:scale-distances}).

For comparison, we also include in Fig.~\ref{fig:Sat_rad_dist} the radial distribution of MW satellites from the Local Volume Database \citep{LG_dwarfs}. The shaded regions represent the $68\%$ confidence interval, obtained by propagating the uncertainties in the reported distances of the MW satellites.

It is clear that the MW satellites are significantly more centrally concentrated than any of the simulated samples considered in this work. 
This discrepancy is a well-known issue  \citep[e.g. ][]{yniguez2014stark,Carlsten2020}
and it seems to be related to the recent infall of the large magellanic cloud \citep[LMC, ][]{patel2024temporal} 
due to both: its gravitational influence, and the group infall of satellite galaxies associated with the LMC.

We do find, for each sample, a large halo-to-halo variance (i.e. variations among individual host galaxies within the same sample) which can account for variations of $\sim40\%$ in the estimation of the half distance (see Table~\ref{tab:scale-distances}).
We indicate such halo-to-halo variations in Fig.~\ref{fig:Sat_rad_dist} with dotted lines. 
Even with this large halo-to-halo scatter, within the inner $100~\kpc$, the MW satellite distribution remains mildly ($\gtrsim1\sigma$) elevated relative to the expected halo-to-halo scatter, but aligns more closely with the \DDO{} sample.

\begin{table}[]   
    \centering
    \caption{Scale measurements of the satellite radial distributions}
    \begin{tabular}{|c|cc|} 
    \hline
    & $\log_{10}{c_{300\kpc}}$ & $d_{\rm half}~[\kpc]$ \\ \hline \hline
    \BDO{} & \rNFWFitLogc{\BulgeDomniatedEnum{}}  &  $165_{-71}^{+33}$ \\
    \CS{} & \rNFWFitLogc{\ControlSampleEnum{}} & $146_{-58}^{+50}$\\
    \DDO{} & \rNFWFitLogc{\DiskDominatedEnum{}} & $118_{-45}^{+50}$ \\ \hline
    MW & $1.2\pm0.5$ & $(49, 84)$ \\ 
    \hline
    \end{tabular}
    \tablefoot{From left to right: base $10$ log concentration in a $300~\kpc$ aperture and the distance enclosing half of the satellites. 
    The reported errors in the two columns are of different nature: for the concentration we report the standard error of the fit, while the errors of $d_{\rm half}$ indicate the $68\%$ CI due to halo-to-halo variance, except for the MW, that we simply report the distance of the 2nd and 4th satellites.}
    \label{tab:scale-distances}
\end{table}

Alternatively, we decided to describe the satellite radial distribution with an NFW profile \citep{Navarro1997}, as it  allows for a better comparison with the dark matter halo concentration.
The fit to the satellite distribution was performed by maximising the log-likelihood defined through the normalised differential enclosed mass profile:
\begin{equation}
    \mathcal{L}(\{r_{ij}\}\mid c) = \prod_{j=1}^{N_{\rm hosts}} \prod_{i=1}^{N_{\rm sat}^j} \frac{1}{M_{200}(c)} \frac{dM(r_{ij}, c)}{dr},
\end{equation}
where $M_{200}(c)$ is the total mass enclosed within $R_{200}$ for an NFW profile with concentration $c$, and
\begin{equation}
    \frac{dM(r, c)}{dr} = 4\pi r^2 \rho_{\rm NFW}(r, c),
\end{equation}
with $\rho_{\rm NFW}(r, c)$ being the standard NFW density profile.
For the radial fits, we fixed $R_{200} \equiv R_{300\,\kpc} = 300\,\kpc$ which corresponds to the maximal aperture within which we identify satellite galaxies. The concentration is defined in terms of the radius where the logarithmic slope of the density profile equals $-2$ ($r_{-2}$), such that:  $c\equiv c_{300\,\kpc}=R_{200}/r_{-2}=300\,\kpc/r_{-2}$.

It is worth noting that, while the NFW fit broadly captures the general shape of the radial distribution of satellites, in the \DDO{} sample it systematically deviates at both small ($r \lesssim 20\,\mathrm{kpc}$) and large radii ($r \gtrsim 150\,\mathrm{kpc}$). 
Specifically, the NFW model overpredicts the number of satellites in the central regions and underpredicts it in the outskirts.
This indicates that satellites in the \DDO{} population are slightly less centrally concentrated than a standard NFW profile would suggest.
However, these deviations are minimal, amounting to fewer than $5\%$.

A direct fit to the satellite radial distributions (assuming an NFW distribution) yields a concentration of $\log_{10}c_{300\kpc} = \rNFWFitLogc{\DiskDominatedEnum{}}$ for the \DDO{} sample, and $\log_{10} c_{300\kpc} = \rNFWFitLogc{\ControlSampleEnum{}}$ for the \CS{} sample. 
When compared to the concentration of the host galaxy’s dark matter halo, these values are lower by about $\sim0.5~{\rm dex}$, a difference similar to the $\sim0.3~{\rm dex}$ offset reported by \citet{McDonough2022} using the \textsc{TNG100} simulation.

This difference in the spatial concentration of satellites does not appear to mirror the concentration of their host dark matter halos. The average halo concentrations\footnote{DM halo concentrations are taken from the public data release of \citep{Anbajagane2022}, where they were estimated via an NFW fit to the dark matter density profile of each subhalo.} are nearly identical for the \DDO{} and \CS{} samples, with values of $\log_{10} c_{200,c}^{\rm DM} = 1.11 \pm 0.01$ and $\log_{10} c_{200,c}^{\rm DM} = 1.10 \pm 0.02$, respectively. Instead, the difference seems to reflect variations in the angular momentum of the host systems. This applies to both the baryonic component, manifested in their different morphologies, and the dark matter component, with the average spin parameter\footnote{Defined as in \cite{Bullock2001}: $\lambda_{\rm DM} = j_{200} / (\sqrt{2} V_{200} R_{200})$, where $j_{200}$ is the total specific angular momentum, $V_{\rm 200}$ is the virial velocity, and $R_{\rm 200}$ is the virial radius.} $\lambda_{\rm DM} $ being roughly $0.2~{\rm dex}$ lower in the \CS{} sample than in the \DDO{} sample (see Appendix \ref{apx:halo-prop}).
This is consistent with the results of \cite{Yetli}, who, in a study of \BDO{} galaxies, found that such systems preferentially undergo mergers that coherently contribute to their angular momentum. We explore the angular distribution of the satellite systems in the following sections.

\subsection{Orbital alignment}
\label{sec:orb-align}

\begin{figure}
    \centering
    \includegraphics[width=1\linewidth]{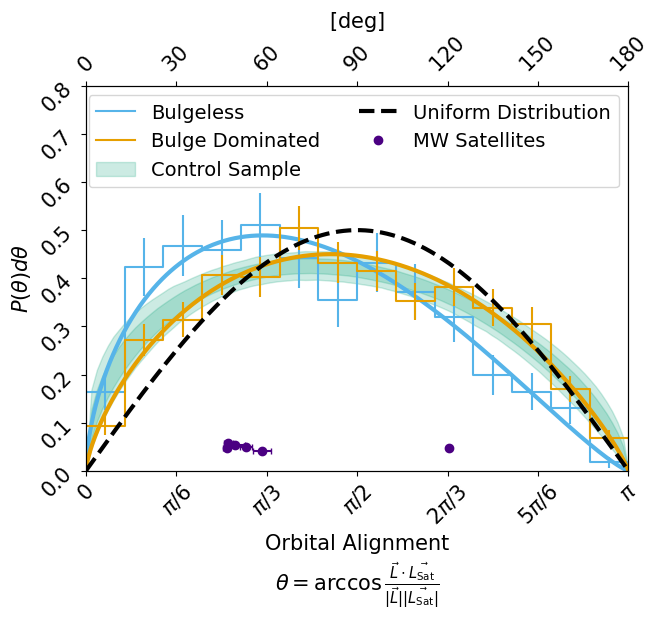}
    \caption{Distribution of satellite orbital alignment with respect to host stellar angular momentum computed within $1$ half mass radius at $z=0$.
    \DDO{} in blue, \BDO{} in orange, and \CS{} in green. 
    Error bars in the histogram indicate $1\sigma$ Poisson scatter.
    Solid thick lines show a best fit to a beta distribution. 
    Green band indicates the $1\sigma$ and $2\sigma$ interval of the best fitted beta distribution to the $1000$ realisations of the \CS{}. 
    The black dashed line indicates what we would expect from an isotropic distribution of satellite orbits 
    We include for comparison the orbital orientation of MW satellites (indigo dots) obtained from the \cite{LG_dwarfs} dataset. Note that for visual purposes we have added random vertical offset to each satellite. }
    \label{fig:sat_ang-dist}
\end{figure}

In Fig.~\ref{fig:sat_ang-dist} we explore the angular distribution of satellite orbits.
To quantify the degree of alignment, we computed the angle between each satellite orbital angular momentum ($\vec{L}_{\rm Sat}$) and the host galaxy stellar angular momentum vector ($\vec{L}$), measured within one stellar half-mass radius: 
\begin{equation}
    \theta =\arccos
    \frac{
        \vec{L}\cdot\vec{L}_{\rm Sat}
        }{
        |\vec{L}||\vec{L}_{\rm Sat}|
        }.
\end{equation}

The dashed black line in Fig.~\ref{fig:sat_ang-dist} represents the expectation for a completely isotropic angular distribution. We find that satellites in the \DDO{} sample exhibit a clear excess at small angles, indicating preferential alignment with the host galaxy angular momentum. This suggests that satellites in \DDO{} systems tend to orbit in a flattened (but thick), co-rotating structure, a planar configuration aligned with the disc plane. 

As a reference, we include in Fig.~\ref{fig:sat_ang-dist} the orbital alignments of the Milky Way (MW) satellites. The strong clustering of the MW satellites reflects the well-known plane of satellites identified in the MW system \cite{}. 
Interestingly, the orientation of the MW satellite plane ($61.0\pm1.5~\deg$ in our selected sub-sample of MW satellites) approximately coincides with the most likely alignment in the \DDO{} sample\footnote{The galaxy near $120~\deg$ is Sculptor, which is counter-rotating but still lies within the same satellite plane.}.
However, we emphasise that connecting the distribution of orbital alignments to individual planes of satellites is not straightforward and requires a dedicated, case-by-case analysis of plane identification, which is beyond the scope of this work and deferred to future study.

This angular alignment complements the findings from the radial distribution (Sec.~\ref{sec:radial}), where \DDO{} galaxies showed more concentrated satellite distributions. Together, these results suggest that the satellite systems of \DDO{} hosts have dynamically colder and more coherent orbital structures than those of the \CS{} or \BDO{} samples.

\subsection{Satellite infall times and alignment evolution}
\label{sec:evolution}

We estimate the satellite infall time as the time when the distance between the satellite and the host galaxy is smaller than the $R_{200}(z)$ of its host. 
Since we use a fixed physical aperture at $z=0$ to identify satellites, in some cases the satellite candidate has never crossed $R_{200}(z)$. 
We call these \emph{pre-infall satellites}.

The fraction of  \emph{pre-infall} satellites is: $1.6\pm 0.6\%$, $0.9\pm0.2$ and $2.3\pm0.8\% $ in the \DDO{}, \BDO{}, and \CS{}, respectively.
The values for the \DDO{} and \CS{} populations are consistent within $1\sigma$, but the \BDO{} sample shows a noticeably smaller fraction.
This  reduction in the \BDO{} sample is primarily due to an aperture selection bias driven by the host-mass distribution: hosts in \BDO{} are more massive and have larger $R_{200}$, so our fixed physical aperture at $z=0$ contains fewer galaxies that lie outside the host $R_{200}$.

\begin{figure}
    \centering
    \includegraphics[width=1\linewidth]{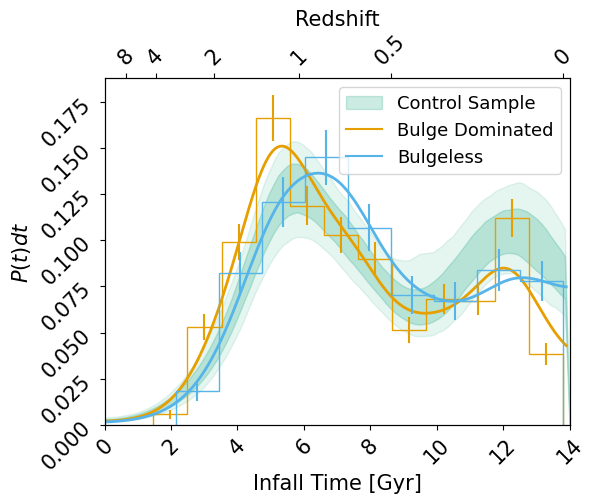}
    \caption{Satellite infall time distribution. 
    Infall times are estimated as the first moment the satellite lies within $R_{200}$ of  its host.
    \DDO{} in blue, \BDO{} in orange, and \CS{} in green. 
    Histogram error bars indicate $1\sigma$ Poisson scatter.
    Solid lines are KDE estimates of their respective distributions.
    Shaded bands in the \CS{} indicate to the $1\sigma$ and $2\sigma$ regions over the $1000$ realisations.  }
    \label{fig:InfallTimes}
\end{figure}

The three populations under study show qualitatively similar distributions of infall times (Fig.~\ref{fig:InfallTimes}). All exhibit a bimodal distribution, with one population of long-lived satellites that had its infall approximately at $z \sim 1$, and a second population of satellites that have infallen recently (at cosmic times $>9-10\,\Gyr$) or are on their first approach.
This bimodal distribution is expected (see \citealt{Simpson2018}) and is due to a selection effect. If we took into account splash-back and already merged galaxies, this bimodality would be expected to vanish.

The long-lived population of satellites infalls slightly later in the \DDO{} sample. There is also an enhancement in the recently infallen satellites in the \CS{} sample. 

\begin{figure}[h!]
    \centering
    \includegraphics[width=1\linewidth]{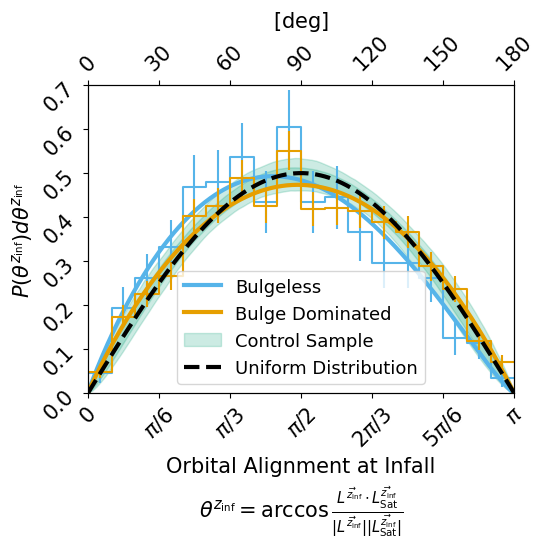}
    \includegraphics[width=1\linewidth]{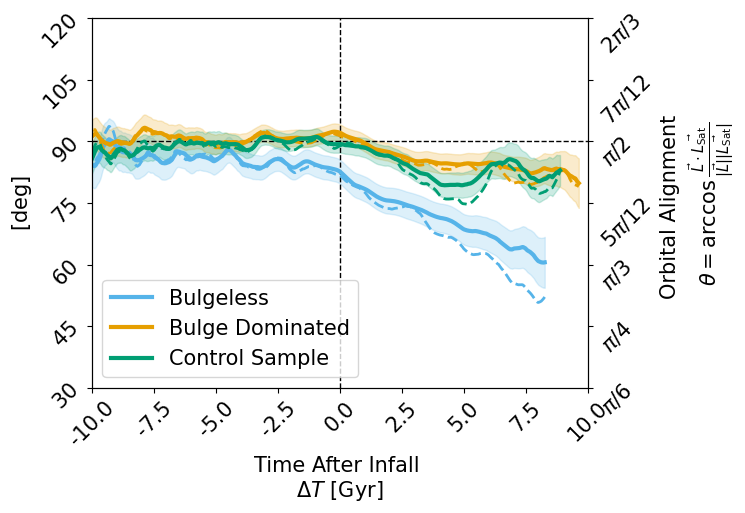}
    \caption{
    Top panel: 
    Distribution of satellite orbital alignment, at infall time, with respect to host stellar angular momentum computed within $1$ half mass radius.
    \DDO{} in blue, \BDO{} in orange and \CS{} in green. 
    Error bars in the histogram indicate $1\sigma$ Poisson scatter.
    Solid thick lines show the best fitting beta distribution. 
    Green band indicates the $1\sigma$ and $2\sigma$ interval of the best fitted beta distribution to the $1000$ realisations of the \CS{}.
    Black dashed line indicates what we would expect from an isotropic distribution of satellite orbits.
    Bottom panel:    
    Evolution of the mean (median) orbital alignment relative to the satellite infall time as a solid (dashed) line. 
    We only show bins with more than $100$ satellites.
    The shaded band indicates the $1\sigma$ uncertainty estimated via bootstrap resampling. 
    Black horizontal thin dashed line at $90\,\deg$ indicates the expected value of an isotropic distribution.
    }
    \label{fig:infal_al-evo}
\end{figure}

We repeated the same experiment as in Sec.~\ref{sec:orb-align} and we explored the orbital alignment of the satellites but at infall time (Fig.~\ref{fig:infal_al-evo} top panel). 
Interestingly, despite the observed angular alignment of satellites in the \DDO{} population at $z=0$, we do not find strong evidence of an anisotropic accretion of the satellite galaxies. 
This is in line with previous studies that found that anisotropic accretion is less prominent in less massive ($M_{\rm halo}\lesssim 10^{13}\,\Msun$) halos \citep[e.g. ][]{Tenneti2021}.

Our results suggest that the coherent satellite configuration observed in \DDO{} hosts may instead arise from long-term orbital evolution within the host potential. We show this in the bottom panel of Fig.~\ref{fig:infal_al-evo}, where we follow the average orbital orientation of the satellites since their infall time, estimated as:
\begin{equation}
    \left<\theta\right>(\Delta t) = {\rm arg}\left(\frac{1}{N_{\rm sat}}\sum_{j=1}^{N_{\rm sat}}\exp{\left(i\theta_{j}(\Delta t)\right)}\right),
\end{equation}
where the summation is done over all the satellites of the sample, $i$ is the imaginary unit, $\theta_{j}$ is the orbital alignment of the $j$-th satellite at a time $\Delta t$ after its infall.
The ${\rm arg}$ function extracts the angle of the resulting complex number. 
The uncertainties were estimated via bootstrap resampling.

As shown in Fig.~\ref{fig:infal_al-evo}, before infall the orbits are, on average, at $90\,\deg$ with respect to the host angular momentum, reflecting the isotropy of the distribution. 
There is a small systematic deviation (before infall) in the \DDO{} sample of the order of $\sim5\,\deg$ which likely reflects the larger angular momentum environment this massive disc galaxies live in \citep[see][]{DiCintio2019,Yetli}.

After infall, the satellite orbits of all three samples tend to steadily align with their host angular momentum at a similar rate: $-2.5\pm0.2\,\deg\,\Gyr{}^{-1}$ (\DDO{}), $-2.1\pm0.13\deg\,\Gyr{}^{-1}$ (\BDO{}) and $-2.36\pm0.17\,\deg\,\Gyr{}^{-1}$ (\CS{}).
This persistent deviation from isotropy is primarily driven by satellites gradually aligning their orbits with the principal axes of their hosts. 
Such behaviour is consistent with classical studies of satellite–host interactions, where tidal torques, dynamical friction, and resonant angular-momentum exchange drive orbital decay, and inclination changes, progressively coupling satellite orbits to the symmetry plane of the central potential \citep[e.g.][]{Quinn1986,Penarrubia2004,Read2008,Ogiya2016}.

At later times (several gigayears after infall), the \BDO{} and \CS{} samples depart from the initial steady alignment: in the \BDO{} sample the alignment rate weakens down to $ -0.64\pm0.12\,\deg\,\Gyr{}^{-1}$, whereas in the \CS{} sample the process effectively stalls or even reverses, reaching $ 0.9\pm0.2\,\deg\,\Gyr{}^{-1}$. In contrast, the \DDO{} satellites maintain their initial alignment rate over the full interval with an alignment rate of $-2.5\pm0.3\,\deg\,\Gyr^{-1}$.

We interpret this divergence as evidence that, although all systems undergo a common secular alignment after infall, this evolution is still susceptible to dynamical perturbations. 
The infall of massive companions, more frequent in \BDO{} and \CS{} hosts, can disturb the coherence of the satellite system and temporarily weaken or stall the alignment, whereas the more weakly perturbed \DDO{} hosts preserve the steady trend over several gigayears. 
Massive accretion events therefore act to modulate an intrinsically morphology-dependent secular process rather than fully determine the observed differences. We assess this interpretation explicitly in Appendix~\ref{apx:orb-by-mass}, where we separate hosts according to the presence or absence of massive infall and compare their alignment histories.

\section{Summary and conclusions}
\label{sec:summary}

In this study we investigated the satellite populations of \DDO{}  galaxies using the high-resolution \TNG{} cosmological simulation. 
Our analysis focused on \DDO{} systems that are morphological analogues of the Milky Way and the primary targets of the \BEARD{} multi-facility survey.
Within \TNG{} simulations we identified MW and \BEARD-like analogues by selecting central galaxies with stellar mass between \StellarMassLowLim{} and \StellarMassUpLim{} that fall in the first quartile of the \BtoD{} mass-ratio distribution. 
For comparison, we also constructed two reference samples: a \BDO{} sample drawn from the upper quartile of the \BtoD{} distribution, and a stellar-mass–matched bulge dominated \CS{} (Fig.~\ref{fig:BD}).  
By systematically comparing the satellite populations of \DDO{}  with both \BDO{} and \CS{}, we uncovered a set of robust trends:

\begin{itemize}
    \item Satellite abundance: We do not find a statistically significant dependence of the number of satellite galaxies with bulge mass fraction. The difference of satellite number counts between the \BDO{} and \DDO{} is due to differences in the mass distribution (Fig.~\ref{fig:nsat-mstar}). 
    \item  Luminosity functions and stellar mass: The satellite luminosity function of \DDO{} hosts displays a significantly ($2.8\sigma$) steeper faint-end slope compared to the \CS{}, indicating an overabundance of faint satellites in \DDO{} galaxies (Fig.~\ref{fig:lum-func}) and  an excess of more massive satellites in the \CS{} (Fig.~\ref{fig:max-mass}).
    Satellite luminosity functions of \DDO{} hosts are in better agreement ($<1\sigma$) with current MW estimations, compared to the \CS{} satellite galaxies ($\sim1.5\sigma$). 
    \item Spatial distribution:
    Satellites of \DDO{} galaxies are more centrally concentrated than those around \BDO{} and \CS{} hosts, with $d_{\rm half} \sim 120~\kpc$ and $\log_{10} c_{300\,\kpc} \approx 0.8$. Despite this, there is substantial halo-to-halo variance, which can change $d_{\rm half}$ by roughly forty per cent. When compared to the Milky Way, whose satellites have $d_{\rm half} = 49$--$84~\kpc$ and $\log_{10} c_{300\,\kpc} \approx 1.2$, all simulated samples remain less centrally concentrated, although the \DDO{} population provides the closest match.
   \item Orbital alignment: A preferential alignment between the orbital angular momentum of satellites and the stellar angular momentum of the host galaxy is found in \DDO{} systems, indicating a preference for co-rotating satellite configurations, in contrast to the more isotropic orbital configurations seen in \BDO{} and \CS{} hosts (Fig.~\ref{fig:sat_ang-dist}). Satellite orbital alignment follows a secular post-infall evolution that tends to align satellite orbits with the host disc (Fig.~\ref{fig:infal_al-evo}). 
   At later times, the alignment weakens or stalls in \CS{} hosts, and to a lesser extent in \BDO{} hosts, while remaining steady in \DDO{} systems. Our extended analysis (Appendix~\ref{apx:orb-by-mass}) indicates that the infall of massive satellite companions can perturb this secular evolution, contributing to the observed slowdown of the alignment in affected systems. In \DDO{} hosts, where massive satellites are rare, the secular alignment proceeds uninterrupted over several gigayears. These results indicate that the efficiency of orbital alignment is intrinsically morphology-dependent, with massive accretion acting to modulate rather than fully determine the trends.
\end{itemize}

Overall, our findings suggest that the morphological state of a galaxy, particularly the absence of a bulge, is closely linked to the properties of its satellite system. 
The steeper luminosity functions, enhanced angular alignment, and spatial concentration of satellites in \DDO{} hosts point to a more dynamically coherent and possibly quiescent evolutionary history (in agreement with the recent work of \citealt{Yetli}). 

These results provide a useful theoretical benchmark for future and current observational campaigns (e.g. \BEARD{}, ELVES, SAGA), and highlight the potential of satellite populations as tracers of the formation history and internal structure of their host galaxies.

\begin{acknowledgements}
We thank the anonymous referee for their constructive comments, which improved the clarity of the paper. 
SCB and JMA acknowledges the support of the Agencia Estatal de Investigación del Ministerio de Ciencia e Innovación (MCIN/AEI/10.13039/501100011033) under grant nos. PID2021-128131NB-I00 and CNS2022-135482 and the European Regional Development Fund (ERDF) ‘A way of making Europe’ and the ‘NextGenerationEU/PRTR’.

AdLC  acknowledges financial support from the Spanish Ministry of Science and Innovation (MICINN) through RYC2022-035838-I and PID2021-128131NB-I00 (CoBEARD project). 

EAG acknowledges support from the Agencia Espacial de Investigación del Ministerio de Ciencia e Innovación (AEI-MICIN) and the European Social Fund (ESF+) through a FPI grant PRE2020-096361.

MCC acknowledges the support of AC3, a project funded by the European Union's Horizon Europe Research and Innovation programme under grant agreement No 101093129. MCC acknowledges financial support from the Spanish Ministerio de Ciencia, Innovación y Universidades (MCIU) under the grant PID2021-123417OB-I00 and PID2022-138621NB-I00.

SZ acknowledges the financial support provided by the Governments of Spain and Aragón through their general budgets and the Fondo de Inversiones de Teruel, the Aragonese Government through the Research Group E16\_23R, and the Spanish Ministry of Science and Innovation and the European Union - NextGenerationEU through the Recovery and Resilience Facility project ICTS-MRR-2021-03-CEFCA.

NCC acknowledges a support grant from the Joint Committee ESO-Government of Chile (ORP 028/2020) and the support from FONDECYT Postdoctorado 2024, project number 3240528. 

EMC and AP acknowledge the support by the Italian Ministry for Education University and Research (MUR) grant PRIN 2022 2022383WFT “SUNRISE” (CUP C53D23000850006) and Padua University grants DOR 2022-2024. 

FP acknowledges support from the Horizon Europe research and innovation programme under the Maria Skłodowska-Curie grant “TraNSLate” No 101108180, and from the Agencia Estatal de Investigación del Ministerio de Ciencia e Innovación (MCIN/AEI/10.13039/501100011033) under grant (PID2021-128131NB-I00) and the European Regional Development Fund (ERDF) ``A way of making Europe’'. 

ADC kindly thanks the Spanish Ministerio de Ciencia, Innovacion y Univerasidades through grant CNS2023-144669, programa Consolidacion Investigadora. 

J.R. acknowledges financial support from the Spanish Ministry of Science and Innovation through the project PID2022-138896NB-C55 and from the Plan Propio de Investigación 2025 submodalidad 2.3 of the University of Córdoba.

The author(s) wish to acknowledge the contribution of the IAC High-Performance Computing support team and hardware facilities to the results of this research. 

This work has extensively used the following software: \texttt{numpy} \citep{numpy}, \texttt{matplotlib} \citep{matplotlib}, \texttt{scipy} \citep{scipy}, \texttt{emcee} \citep{emcee}, \texttt{ArviZ} \citep{arviz}, \texttt{bambi} \citep{bambi}, \texttt{pynbody} \citep{pynbody} and \texttt{numba} \citep{numba}.

\end{acknowledgements}

\bibliographystyle{aa}
\bibliography{biblio}

\appendix

\section{Halo mass distribution}
\label{apx:halomass}

\begin{figure}[!hb]
    \centering
    \includegraphics[width=1\linewidth]{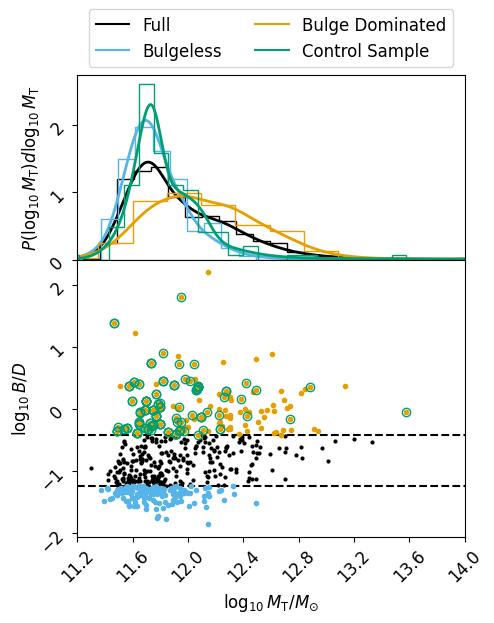}
    \caption{
    Top panel: Distribution of host galaxy stellar mass. 
    Bottom panel: Bulge-to-disc (B/D) mass ratio of host galaxies as a function of total halo mass. \\
    Both panels display the full parent sample (black), the \DDO{} sample (blue), the \BDO{} sample (orange), and, for clarity, one realisation of the \CS{} sample (green).
    In the bottom panel, black horizontal dashed lines mark the first and fourth quantiles of the \BtoD{} distribution, which define the division between \DDO{} and \BDO{} galaxies.
    }
    \label{fig:selection-SH}
\end{figure}

As discussed in Section~\ref{sec:number-counts}, halo mass is the primary factor governing both the number of satellites and their properties.
In this section, we therefore present the halo mass distributions of the samples explored in this work (Fig.~\ref{fig:selection-SH}).
\BDO{} galaxies tend to reside in more massive halos, by roughly $\sim0.5~{\rm dex}$, compared to the \DDO{} sample.
However, the halo mass distributions of the galaxies of the \CS{} and \DDO{} samples are statistically indistinguishable.
A Kolmogorov–Smirnov test yields a median p-value of $1.380$, confirming that the two distributions are fully consistent with being drawn from the same parent population.

\section{Mass Accretion Histories}
\label{apx:MAH}

\begin{figure}[h!]
    \centering
    \includegraphics[width=1\linewidth]{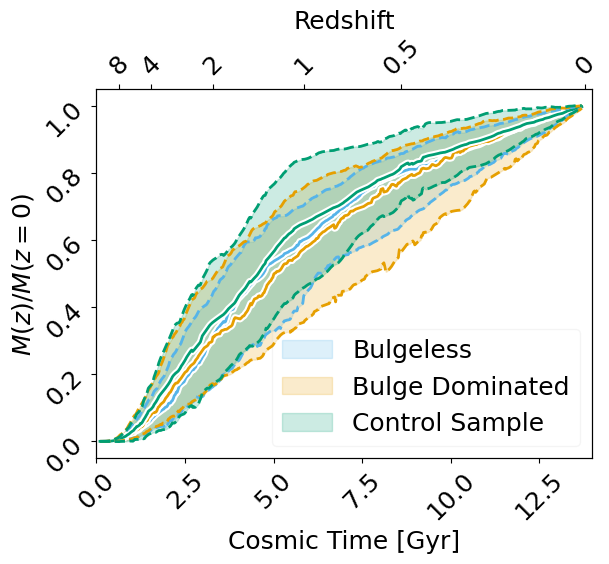}
    \caption{
    Mass accretion histories (baryons plus dark matter), normalised to the final mass. 
    Shaded bands indicate the halo-to-halo scatter, enclosing the 16th and 84th percentiles of the MAHs for each galaxy. 
    Dashed lines are included to delineate the same region and aid visualisation in cases where the shaded bands overlap.
    }
    \label{fig:MAH}
\end{figure}

In this appendix, we present the mass accretion histories (MAHs) of the galaxy samples analysed in this work. 
The MAHs include both baryons and dark matter and are normalised to the final mass of each galaxy (Fig.~\ref{fig:MAH}).

\BDO{} hosts, on average, form slightly more slowly than galaxies from the \DDO{} and \CS{} samples, reaching half of their stellar mass approximately $0.4\,\Gyr$ later (see Table~\ref{tab:t50}). 
Nevertheless, there is substantial variability due to the stochastic nature of individual MAHs, with the $1\sigma$ scatter in $t_{50}$ being $\sim2.3,\Gyr$ for \BDO{}, $\sim 2.5\,\Gyr$ for \CS{}, and $\sim1.9\,\Gyr$ for \DDO{} galaxies.

Although the scatter in MAHs is larger in the \CS{} sample than in the \DDO{} sample, the average evolution of galaxies in these two samples is nearly identical. 
Despite this, the growth of the stellar component (not shown) differs between the samples, with the \CS{} and \BDO{} galaxies exhibiting faster stellar mass growth, in agreement with the findings of \cite{Yetli}.
This indicates that differences in the assembly histories of host galaxies cannot account for the observed variations in the properties of their satellite populations.

\begin{table}[h!]
    \centering
    \caption{ Average half-mass assembly time and standard deviation for \BDO{}, \CS{}, and \DDO{} galaxies.}
    \begin{tabular}{|c|cc|} \hline
        Sample & $<t_{50}>~[\Gyr]$ & $\sigma_{t_{50}}~[\Gyr]$ \\ \hline \hline
        \BDO{} & $5.48$ 
        & $2.3$ \\
        \CS{}  & $5.13 \pm 0.20$ & $2.5$ \\
        \DDO{} & $5.12$ 
        & $1.9$ \\ \hline
    \end{tabular}
    \tablefoot{Uncertainty in the \CS{} indicate the variability across the $1000$ realisations. }
    \label{tab:t50}
\end{table}

\section{Host Halo DM concentrations and spin}
\label{apx:halo-prop}

Dark matter halo concentrations, $c_{200,c}^{\rm DM}$, were obtained from \cite{Anbajagane2022}, where they were estimated by fitting an NFW profile to the dark matter density profile of each subhalo out to $R_{200,c}$, i.e. the radius enclosing a  density $200$ times the critical density of the Universe.

For the spin parameter, we adopt the definition of \cite{Bullock2001}:
\begin{equation}
    \lambda_{\rm dm} = \frac{j_{200,c}}{\sqrt{2} \, V_{200,c} \, R_{200,c}},
\end{equation}
where $j_{200,c}$ is the specific dark matter halo angular momentum, $V_{200,c} = \sqrt{\frac{G M_{200,c}}{R_{200,c}}}$,  $G$ the gravitational constant, $R_{200,c}$ is the virial radius, and $M_{200,c}$ is the dark matter halo mass. For consistency with the concentration definition, all these quantities were measured within an aperture of $R_{200,c}$.

\begin{figure}[h]
    \centering
    \includegraphics[width=1\linewidth]{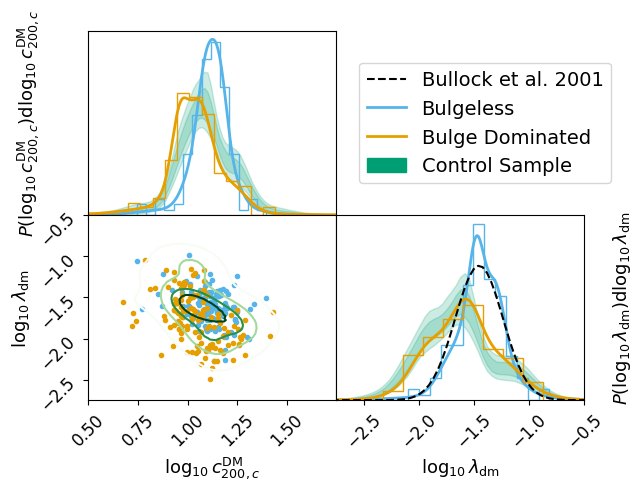}
    \caption{
    Corner plot showing the joint and marginal distributions of dark matter halo concentration, $c_{200,c}^{\rm DM}$, and spin parameter, $\lambda$, for the \DDO{} (blue), \BDO{} (orange) and \CS{} (green). 
    Diagonal panels show the normalised marginal distribution.
    Shaded bands indicate $1\sigma$ and $2\sigma$ confidence regions over the $1000$ realisations of the \CS{}. 
    Concentrations are obtained via an NFW fit to the DM density profile from \cite{Anbajagane2022}
    The $2$D contours indicate the $0.5\sigma$, $1\sigma$, $1.5\sigma$ and $2\sigma$ confidence regions. 
    Black line dashed in bottom right panel indicates the expected spin parameter $\lambda_{\rm dm}$ distribution from \cite{Bullock2001}.
}
    \label{apx:fig:corner}
\end{figure}

Fig.~\ref{apx:fig:corner} shows a corner plot of the distributions and correlations between the dark matter halo concentration, $c_{200,c}^{\rm DM}$, and spin parameter, $\lambda_{\rm dm}$, for the \DDO{}, \BDO{}, and \CS{}. The expected distribution of dark matter halo spin from \cite{Bullock2001} is also indicated for reference.

As expected from the concentration-mass relation \citep{Dutton2014} the \BDO{} sample shows smaller concentrations than both \DDO{} and \CS{}. 

\CS{} and \DDO{} samples have approximately equal average concentrations, with $\log c_{200, c}^{\rm DM}$ $-1.10\pm0.02$ and $-1.114\pm 0.008$ respectively. 
While the average are similar, the scatter is slightly larger in the \CS{}, $0.22\pm0.03$ compared to the $\sim0.097\pm0.009$ found in the \DDO{}.  
The uncertainties in the average and scatter are bootstrap estimates.

Regarding the DM halo spins, there is clear systematic difference between the \DDO{} galaxies and the \CS{}. 
Having the \CS{} (and the \BDO{} galaxies), extremely low values, even below the expectations from \cite{Bullock2001}. 

In summary, DM halos of \DDO{} and \CS{} galaxies are radially similar, but show systematic differences in their angular momentum.

\section{The effect of the massive infall in orbital alignment}
\label{apx:orb-by-mass}

In this section we examine the impact of massive infalling satellites on the evolution of the orbital alignment of satellite systems.
To this end, we analyse three complementary selections:
\begin{itemize}
\item satellites in hosts that experience a massive satellite infall (\BH{});
\item satellites in hosts without such an event (\BL{});
\item a cleaned sample in which the orbital evolution of each satellite is followed only up to the moment when a massive infall occurs in its host, if applicable.
\end{itemize}

If perturbations induced by massive satellites are the dominant driver of alignment changes, we expect the \BL{} sample to follow the intrinsic secular alignment reported in Section~\ref{sec:orb-align}, whereas in the \BH{} sample a slowdown or stalling of the alignment should appear a few gigayears after the infall of the massive companion.
For the cleaned sample, we expect the evolution to track that of the full population until the time when the effects of the massive infall begins to be present in the \BH{} systems.

\begin{figure}
    \centering
    \includegraphics[width=1\linewidth]{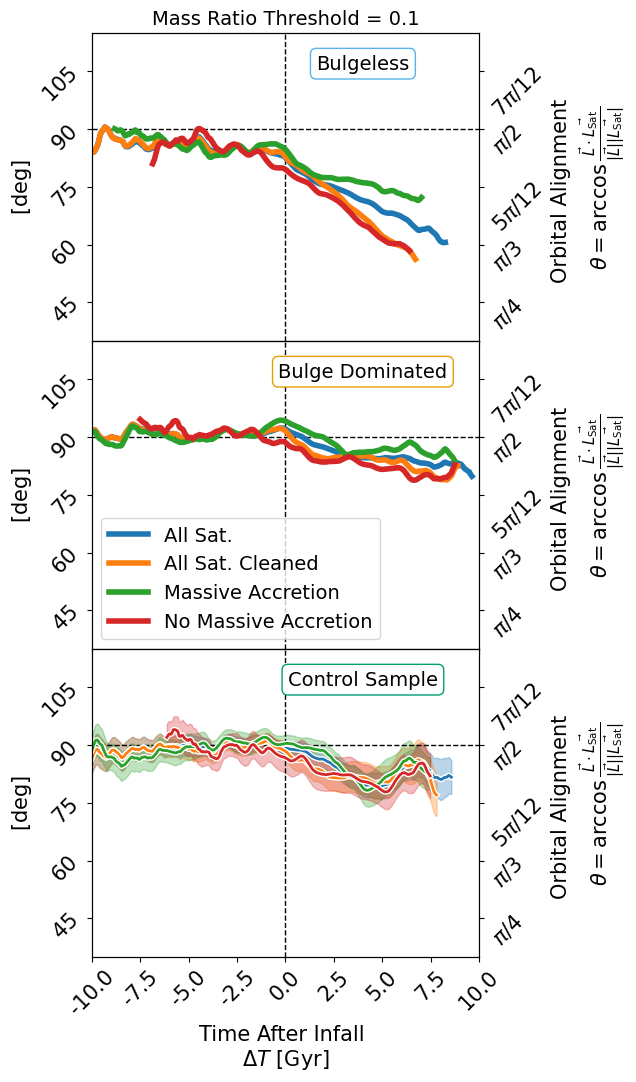}
    \caption{ Mean orbital–alignment evolution for satellites in the \DDO{} (top), \BDO{} (middle), and \CS{} (bottom) samples. Curves correspond to the full sample with all the satellites (blue), satellites in hosts that underwent massive satellite infalls (\BH{}; red), satellites in hosts without a massive infall (\BL{}; green), and the cleaned sample tracking satellite orbits up to the moment of the massive infall if any (orange). Any satellite with a maximum mass ratio with the host galaxy $>0.1$ is considered a massive infall.
    The comparison highlights systematic differences between the subsamples and allows for assessment of the impact of massive satellites on the alignment trends.
    }
    \label{fig:orb-align-mass-thr}
\end{figure}

Figure~\ref{fig:orb-align-mass-thr} shows the alignment evolution of these three selections: \BH{} (green), \BL{} (red), and cleaned (orange); compared to the full sample (blue).
Here, a massive infall is defined as a satellite–to–host mass ratio greater than $>0.1$ at the time when the satellite reaches its maximum total mass.
We explored higher thresholds, which yield qualitatively similar trends, although the statistics of the \BH{} sample naturally decrease as the threshold increases.

Clear systematic differences emerge between the \BH{} and \BL{} populations: satellites in hosts with \BH{} tend to be, on average, less aligned than those in \BL{} hosts.
Interestingly, the divergence between the two samples begins roughly 
$\sim1~\Gyr$ before the infall time, suggesting that the perturbative influence of the massive satellite is already felt by the satellite system prior to the companion crossing the host's virial radius.

The dependence of these effects on host morphology also reveals notable differences.
In \DDO{} galaxies (top panel of Fig.~\ref{fig:orb-align-mass-thr}), the expected behaviour is most clearly visible: both the full sample and the \BH{} sample show a pronounced slowdown in their alignment evolution, whereas this feature is absent in the \BL{} and cleaned samples.

The \BDO{} hosts (middle panel) show a more complex behaviour.
Their average alignment is closer to isotropy (Section~\ref{sec:orb-align}), making the slowdown in the \BH{} sample less prominent, though still present.
Meanwhile, the \BL{} and cleaned samples are almost indistinguishable and lie only a few degrees below the full-sample trend, so the significance of the difference is modest.
A similar qualitative picture applies to the \CS{} sample (bottom panel), but in this case the sample variance is too large relative to the expected signal to draw firm conclusions.

In summary, the differential slowdown associated with the infall of a massive satellite is strong and unambiguous in the \DDO{} hosts, and present but less statistically significant in the \BDO{} hosts due to their lower intrinsic alignment levels.
For the \CS{} sample, the effect remains inconclusive because of the large sampling variance.
These findings also suggest that the underlying mechanism driving the secular alignment differs in strength (or potentially in nature) between \DDO{} and \BDO{} galaxies.

\end{document}